\documentclass[fleqn,usenatbib]{mnras}

\usepackage{newtxtext,newtxmath}

\usepackage[T1]{fontenc}

\DeclareRobustCommand{\VAN}[3]{#2}
\let\VANthebibliography\thebibliography
\def\thebibliography{\DeclareRobustCommand{\VAN}[3]{##3}\VANthebibliography}

\usepackage{graphicx}	
\usepackage{amsmath}	
\usepackage{multirow}
\usepackage{subfig}
\usepackage{xcolor}

\newcommand{\NII}{{[N\,{\sc ii}]}}
\newcommand{\NIIs}{{[N\,{\sc ii}]\,}}

\newcommand{\SII}{{[S\,{\sc ii}]}}

\newcommand{\SIII}{{[S\,{\sc iii}]}}

\newcommand{\OIII}{{[O\,{\sc iii}]}}

\newcommand{\OIIIs}{{[O\,{\sc iii}]\,}}

\newcommand{\OII}{{[O\,{\sc ii}]}}

\newcommand{\OI}{{[O\,{\sc i}]}}

\newcommand{\HeI}{{He\,{\sc i}\,}}

\newcommand{\Ha}{H$\alpha$}
\newcommand{\Has}{H$\alpha$\,}
\newcommand{\Hb}{H$\beta$}

\newcommand{\Hbs}{H$\beta$\,}

\title[The AGN Rosetta Stone]{JADES - The Rosetta Stone of JWST-discovered AGN: deciphering the intriguing nature of early AGN
}

\author[Juodžbalis et al.]{
Ignas Juodžbalis,$^{1, 2}$\thanks{E-mail: ij284@cam.ac.uk}
Xihan Ji,$^{1, 2}$
Roberto Maiolino,$^{1, 2, 3}$
Francesco D'Eugenio,$^{1, 2}$
Jan Scholtz,$^{1, 2}$ \newauthor
Guido Risaliti,$^{4, 5}$ Andrew C. Fabian,$^{6}$ Giovanni Mazzolari,$^{1, 7, 8}$ Roberto Gilli,$^{8}$ Isabella Prandoni,$^{9}$ \newauthor
Santiago Arribas,$^{10}$ Andrew J.\ Bunker,$^{11}$ Stefano Carniani,$^{12}$
Stéphane Charlot,$^{13}$ Emma Curtis-Lake,$^{14}$\newauthor Anna de Graaff,$^{15}$  Kevin Hainline,$^{16}$ Eleonora Parlanti,$^{12}$ Michele Perna,$^{10}$ Pablo G. Pérez-González,$^{10}$\newauthor
Brant Robertson,$^{17}$ Sandro Tacchella,$^{1, 2}$ Hannah \"Ubler,$^{1, 2}$Christina C. Williams,$^{18}$ Chris Willott,$^{19}$ \newauthor Joris Witstok$^{1, 2}$
\\
$^{1}$Kavli Institute for Cosmology, University of Cambridge, Madingley Road, Cambridge, CB3 OHA, UK.\\
$^{2}$Cavendish Laboratory - Astrophysics Group, University of Cambridge,
19 JJ Thomson Avenue, Cambridge, CB3 OHE, UK.\\
$^{3}$ Department of Physics and Astronomy, University College London, Gower Street, London WC1E 6BT, UK \\
$^{4}$Dipartimento di Fisica e Astronomia, Università di Firenze, via G. Sansone 1, 50019 Sesto Fiorentino, Firenze, Italy \\
$^{5}$
INAF – Osservatorio Astrofisico di Arcetri, Largo Enrico Fermi 5, I-50125 Firenze, Italy \\
$^{6}$ Institute of Astronomy, University of Cambridge, Madingley Road, Cambridge CB3 0HA, UK \\
$^{7}$  Dipartimento di Fisica e Astronomia, Università di Bologna, Via Gobetti 93/2, I-40129 Bologna, Italy\\
$^{8}$  INAF – Osservatorio di Astrofisica e Scienza dello Spazio di Bologna, Via Gobetti 93/3, I-40129 Bologna, Italy\\
$^{9}$ INAF – Istituto di Radioastronomia, Via Gobetti 101, I-40129 Bologna, Italy\\
$^{10}$ Centro de Astrobiolog\'ia (CAB), CSIC–INTA, Cra. de Ajalvir Km.~4, 28850- Torrej\'on de Ardoz, Madrid, Spain\\
$^{11}$Department of Physics, University of Oxford, Denys Wilkinson Building, Keble Road, Oxford OX1 3RH, UK\\
$^{12}$Scuola Normale Superiore, Piazza dei Cavalieri 7, I-56126 Pisa, Italy\\
$^{13}$Sorbonne Universit\'e, CNRS, UMR 7095, Institut d'Astrophysique de Paris, 98 bis bd Arago, 75014 Paris, France\\
$^{14}$Centre for Astrophysics Research, Department of Physics, Astronomy and Mathematics, University of Hertfordshire, Hatfield AL10 9AB, UK\\
$^{15}$Max-Planck-Institut f\"ur Astronomie, K\"onigstuhl 17, D-69117, Heidelberg, Germany\\
$^{16}$Steward Observatory, University of Arizona, 933 N. Cherry Avenue, Tucson, AZ 85721, USA\\
$^{17}$Department of Astronomy and Astrophysics University of California, Santa Cruz, 1156 High Street, Santa Cruz CA 96054, USA\\
$^{18}$NSF’s National Optical-Infrared Astronomy Research Laboratory, 950 North Cherry Avenue, Tucson, AZ 85719, USA\\
$^{19}$NRC Herzberg, 5071 West Saanich Rd, Victoria, BC V9E 2E7, Canada
}

\date{Accepted XXX. Received YYY; in original form ZZZ}

\pubyear{2024}

\begin{document}
\label{firstpage}
\pagerange{\pageref{firstpage}--\pageref{lastpage}}
\maketitle

\begin{abstract}
JWST has discovered a large population of Active Galactic Nuclei (AGN) at high redshift, which are weak in the X-rays. Here we present the NIRSpec spectrum of the most extreme of these objects, GN-28074, an AGN at $z=2.26$ with prominent hydrogen and \HeI broad lines, and with the highest limit on the bolometric to X-ray luminosity ratio among all spectroscopically confirmed AGN in GOODS. This source is also characterized by a mid-IR excess, likely associated with the AGN torus' hot dust. The high bolometric luminosity and moderate redshift of this AGN allow us to explore its properties more in depth relative to other JWST-discovered AGN. The NIRSpec spectrum reveals prominent, slightly blueshifted absorption of H$\alpha$, H$\beta$ and \HeI$\lambda$10830. The Balmer absorption lines require gas with densities of $n_{\rm H}> 10^8~{\rm cm}^{-3}$, consistent with clouds in the Broad Line Region (BLR). This finding suggests that part of the X-ray weakness is due to high (Compton thick) X-ray absorption by clouds in the BLR, or in its outer regions.
GN-28074 is also extremely radio-weak. The radio weakness can also be explained in terms of absorption, as the inferred density of the BLR clouds makes them attenuate the radio emission through free-free absorption. Alternatively, the nuclear magnetic field may be underdeveloped, resulting both in intrinsically weak radio emission and lack of hot corona, hence intrinsic X-ray weakness. Finally, we show that recently proposed scenarios, invoking hyper-dense outflows or Raman scattering to explain the broad H$\alpha$, are ruled out.

\end{abstract}

\begin{keywords}
galaxies: active -- quasars: supermassive black holes
-- galaxies: Seyfert
\end{keywords}



\section{Introduction}
The first two years of operation of the James Webb Space Telescope (JWST) have revealed a large population of broad line active galactic nuclei (AGN) at high redshift \citep{Harikane2023,Kocevski2023,Matthee2024,Maiolino2024GNz11,Maiolino2023,Greene2024,Kokorev2023BLAGN,Furtak2024BLAGN,
Wang2024, Wang2024_z7, Ubler2024}. These JWST-selected AGN span the redshift range from $\sim$3 to $\sim$11 and  have bolometric luminosities ranging from $10^{42}$ to $10^{45}$~erg~s$^{-1}$. Therefore, they cover the intermediate/low-luminosity regime of AGN activity inaccessible by pre-JWST high-redshift quasar surveys.

These objects display unambiguous evidence for the presence of an AGN, not only in terms of luminous, broad (FWHM $\sim$ 3000~km~s$^{-1}$) \Has and \Hbs lines (which lack a kinematically similar counterpart in the \OIIIs$\lambda \lambda 4959,5007$ doublet, hence cannot be ascribed to outflow), but also some of them have other transitions typical of AGN, such as high ionization and even coronal lines \citep{Ubler2023,Ji2024GS3073,Ji2024GN-z11,Maiolino2024GNz11}.

However, these AGN discovered by JWST spectroscopy also exhibit unusual properties when compared to standard Type 1 AGN.
A particularly notable property of JWST-selected low-luminosity Type 1 AGN is their X-ray weakness,
which has been pointed out by various authors
\citep{Maiolino_xray_weak,Yue2024,Ananna2024}.
%
These works reported no X-ray detections for the vast majority of Type 1 AGN selected by JWST, even when stacked together. The same X-ray weakness has been found also for narrow line AGN candidates discovered by JWST 
\citep{Scholtz2023AGN2,Chisholm2024AGN2} and for mid-IR selected AGN \citep{Lyu2024}. In addition, a fraction of JWST-selected Type 1 AGN show the presence of absorption in the Balmer lines \citep{Matthee2024, Kocevski2024, Wang2024}. We should mention that examples of X-ray weak AGN with mild optical obscuration, and broad-lined AGN, were already found at high-z even before the advent of JWST data and using ground based surveys \citep[e.g.,][]{Fujimoto2022,Ma2024}.

In the low-redshift Universe, X-ray weakness of type 1 AGN is not common, but it has been observed in a number of metal-poor and low-mass Type 1 AGN candidates that the X-ray emission is weak or lacking \citep[e.g.,][]{simmonds2016,cann2020}.
Among the low-redshift and metal-poor AGN candidates, \citet{burke2021} reported observations of an absorption feature in the broad \Ha\ line in SDSS J102530.29+140207.3 at $z = 0.1$, suggesting large column densities of neutral hydrogen and thus obscuration in the nuclear region.
It is noteworthy that outside the metal-poor AGN sample, there have also been detections of strong broad Balmer-line absorptions in general Type 1 AGN up to a redshift of $z\sim 2$ in the pre-JWST era, but at a very low detection rate \citep[e.g.,][]{hutchings2002,aoki2006,aoki2010,hall2007,ji2012,ji2013,wangxu2015,Zhang2015,Shi2016,williams2017,Schulze2018,hamann2019}.
With typical values of $10^5-10^7~{\rm M_{\odot}}$, the black hole masses of the local metal-poor AGN candidates are comparable to the Type 1 AGN at early times currently revealed by JWST.

Several explanations of this X-ray weakness in JWST-confirmed AGN have been proposed thus far.
One possibility is that (hard) X-rays are absorbed by a Compton thick medium ($N_{\rm H}>10^{24}~{\rm cm}^{-2}$). However, given that X-ray weakness is seen also for Type 1, broad line AGN, the absorbing medium should have comparatively little dust content as dust sufficient to absorb X-rays would also absorb the BLR emission, assuming typical dust-to gas ratios. One possibility that has been proposed is that the BLR clouds could be a potential absorbing medium, as they are expected to be dust free and with large column of gas. However, given that the X-ray weakness affects most JWST-selected AGN, the covering factor of the BLR clouds in these objects should be much higher than in classical AGN (which is typically of $C_f \equiv \Omega/4\uppi \sim 0.5$ at low Eddington ratios; \citealp{Ferland2020}); this scenario seems to be confirmed by the larger equivalent width of the broad H$\alpha$ in JWST-selected AGN relative to classical AGN \citep{Maiolino_xray_weak,Wang2024Rubies-z7}.
Alternatively, the X-ray weakness might be intrinsic to this population of AGN, possibly due to a lack of a corona. JWST-selected AGN tend to share properties similar to those of the so-called Narrow Line Sy1, characterized by a steep X-ray spectrum \citep{Vasudevan2007,Tortosa2022,Tortosa2023} and some of them are characterised by high accretion rates, which typically results in a steep X-ray spectrum \citep{Hagen2024}. The steepened X-ray spectra would thus make the detection of hard X-ray emission in the highly accreting high-$z$ AGN very difficult.
 
However, further verification of these scenarios is difficult, because without X-ray detections there is no simple method of constraining the Compton thickness or the shape of the X-ray continuum (although the very few X-ray detected and X-ray weak JWST-selected AGN appear heavily absorbed; \citealp{Maiolino_xray_weak}). However, one possible way of exploring heavy X-ray absorption and, possibly, Compton thickness would be via the detection of certain absorption lines that could be used to provide estimates on the total hydrogen column density obscuring the source.

On the other hand, the radio properties of the X-ray weak AGN discovered by JWST have yet to be explored in detail. However, radio emission can provide crucial insights into the origin of the observed X-ray weakness, given the well-established connection between AGN X-ray and radio luminosities reported in several studies \citep[e.g.][for a review]{Merloni2003, Panessa2019, Bariuan2022radio, Wang2024radio}. In the local universe, analyses of radio emission from X-ray weak sources have sometimes revealed corresponding radio weakness \citep{Johnson2009SBS, Arcodia2024}, while in other cases, the radio emission is found at the expected levels \citep{Berton2018radio, PaulBrabndt2024}, with detections of compact radio cores or small-scale jets \citep{Meyer2024_jet, Singha2023, Su2021}. A more detailed discussion of the radio properties of JWST-detected BLAGN will be presented in a forthcoming paper (Mazzolari et al., in prep).

In this paper we present the analysis of GN-28074 (RA = 189.06458, Dec = 62.27382; CANDELS ID GN-19549; 3D-HST ID 26945) a broad line AGN at $z = 2.26$ detected by JWST in the GOODS-N field and exhibiting clear absorption features in the \Ha, \Hbs and \HeI$\lambda$10830 lines. As we discuss in this paper, the presence of these lines reveals the existence of gas with moderately high temperature ($\sim$ 10,000~K) and high density and high column densities, likely associated with the BLR or its outer parts. In addition, the continuum of the source exhibits similar colors to the recently identified red sub-population known as the Little Red Dots (LRDs); \citep{Labbe2023, Furtak2023, Kokorev2024, Matthee2024}, with it passing most the color cuts presented in \cite{Kokorev2024} if redshifted to $z = 7$ (F277W - F356W = 0.96, F277W - F444W = 1.64, F150W - F200W = 0.88). An additional component of dust emission at long wavelengths further confirms the presence of a reddened AGN or a dusty torus. However, despite the unambiguous presence of an AGN, this source is undetected both in the X-rays \citep{Maiolino_xray_weak} and in the radio (Mazzolari in prep.). We will show that, therefore, this AGN is a unique case study to understand the properties of the new class of AGN that is being discovered by JWST.

The text is organized as follows.  In \autoref{sec:reduction} we present a summary of the spectroscopic data reduction. \autoref{sec:fitting} contains a description of our line fitting procedure. The bulk of the analysis, pertaining to the detected absorption lines, properties of the central black hole and the continuum shape is presented in \autoref{sec:analysis}. In the following \autoref{subsec:alt_blr} we briefly explore alternative scenarios proposed to explain broad emission in JWST sources. Discussion on potential local analogues and BLR absorption in other spectroscopic JWST AGN is presented in \autoref{subsec:analogues} and \autoref{subsec:blr_absorption} respectively. Finally, in \autoref{sec:conclusion}, we summarize our findings and put them in context of the previous observations. Throughout the paper, we assume flat $\Lambda$CDM cosmology with $\Omega_m = 0.315$ and $H_0 = 67.4$ km~s$^{-1}$~Mpc$^{-1}$.

\section{Data reduction}
\label{sec:reduction}
\begin{figure*}
   \includegraphics[width=\textwidth]{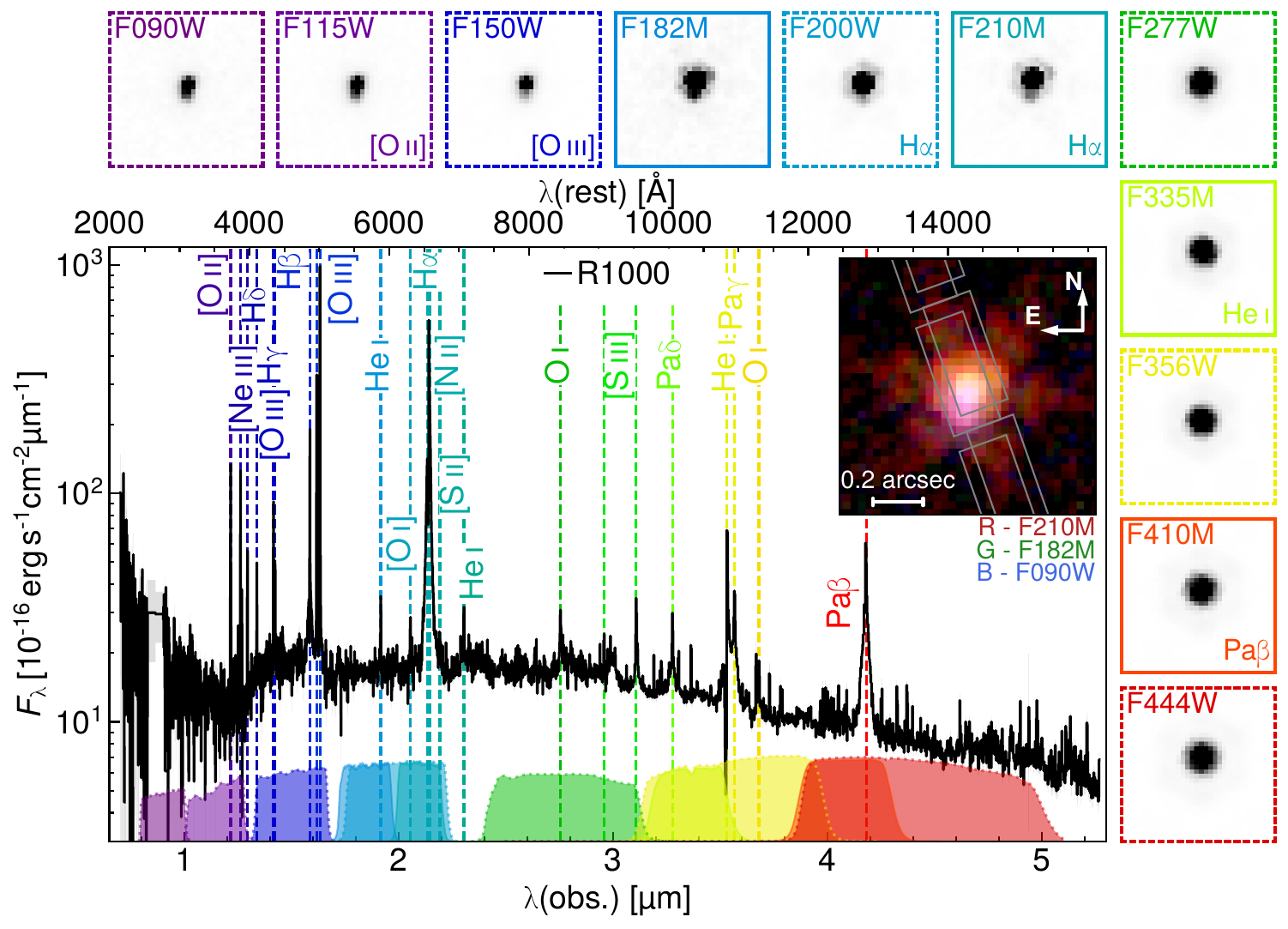}
    \caption{R1000 spectrum of GN-28074 with emission lines marked. The fluxes on the y axis are presented in {\it logarithmic} scale to better showcase weaker lines and the shape of the continuum. All bright permitted lines appear to have broad components while the continuum showcases the characteristic `v' shape found in some objects of the LRD population. The false-colour NIRCam image (inset) highlights the point-source nature of the object, and the position of the MSA slitlets (grey rectangles). We note a pink region to the south-east, denoting a deficiency in F182M flux (see text for a discussion). The rest of the NIRCam single-band cutouts highlight the extended emission (only discernible at SW wavelengths); the angular size of these cutouts is the same as the RGB image.}
    \label{fig:full_spec}
\end{figure*}

Data utilized in this paper has been obtained as part of the JWST Advanced Extragalactic Survey (JADES) program \citep{JADES_desc}. The spectroscopic observations are split into tiers depending on whether they were selected with HST or JWST instruments and depth of exposure. Spectroscopy of our source was obtained as part of the Medium/HST tier in the GOODS-N field and consisted of 1.7 hours of exposure per source in the prism and 1.7 hours per source in each of the medium-resolution gratings, G140M/F070LP, G235M/F170LP and
G395M/F290LP. The observations used two dithered positions (grey rectangles in the inset figure of \autoref{fig:full_spec}), with three nods each.

Data reduction procedures were largely similar to other JADES papers \citep{Bunker2023, Carniani2023, Deugenio2024}. In summary, the pipeline used was developed by
the NIRSpec GTO team and the ESA NIRSpec Science Operations Team. As the absorption features appear to be unresolved, we employ 1D spectra extracted from the central 3 pixels (at 0.1$''$ per pixel scale) of the 2D spectra.

An additional post-processing procedure was carried out after the main data reduction stage as the sigma clipping employed by the GTO pipeline has erroneously flagged strong emission lines and rapid changes in emission due to deep absorption features. This procedure consisted of utilizing the unclipped spectrum to visually identify incorrectly clipped regions and substitute them into the clipped data. The final spectrum obtained by combining the three grating spectra is shown in \autoref{fig:full_spec}.

We also use the multi-band photometry from the NIRCam component of the JADES survey \citep{Rieke2023} and from FRESCO \citep[First Reionization Epoch Spectroscopically Complete Observations;][]{Oesch2023}, combined with HST and Spitzer photometry from the Rainbow database\footnote{https://arcoirix.cab.inta-csic.es/Rainbow\_Database/Home.html}.

\section{Emission and absorption lines fitting}
\label{sec:fitting}
\begin{figure*}
    \centering
    \subfloat[\Hbs + \OIII]{\includegraphics[width=1\columnwidth]{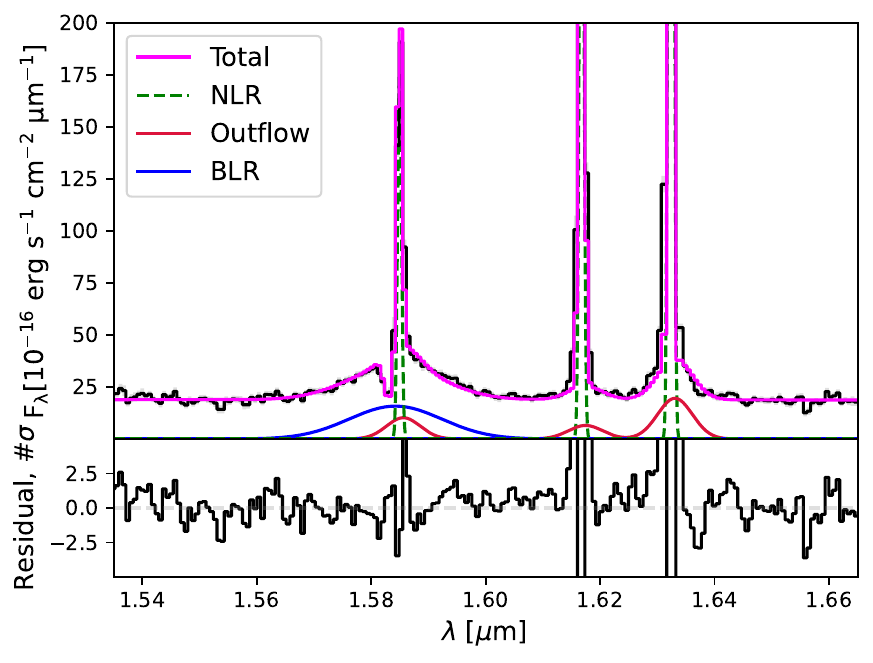}} \hfill
    \subfloat[\Has + \NII]{\includegraphics[width=1\columnwidth]{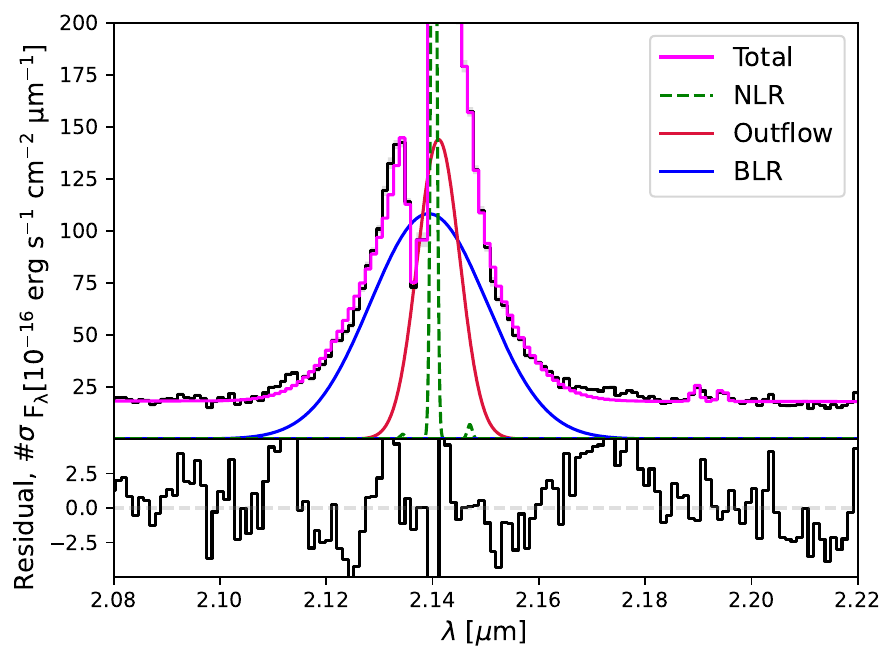}} \hfill
    \subfloat[\HeI\ + Pa$\gamma$]{\includegraphics[width=1\columnwidth]{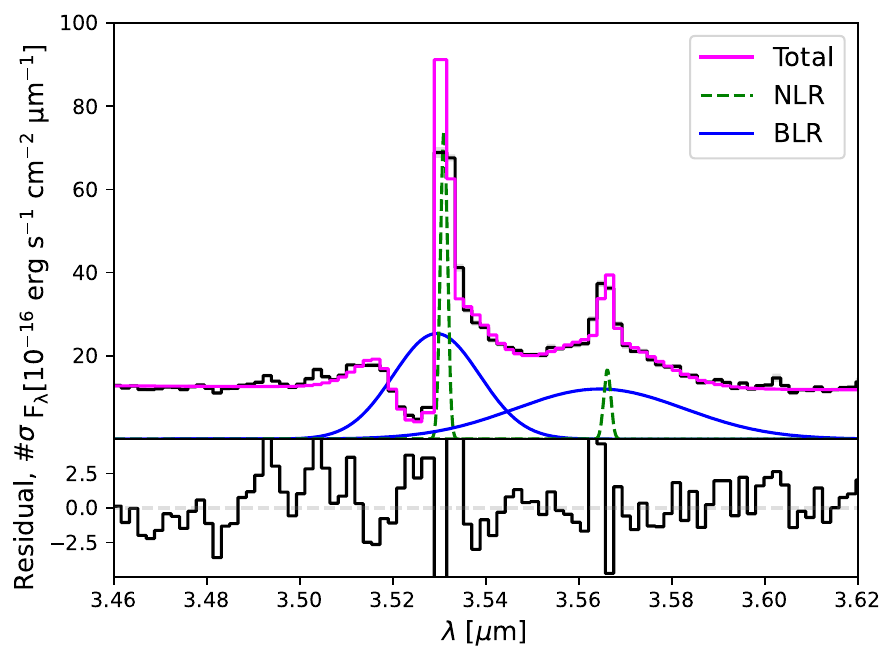}} \hfill
    \subfloat[Pa$\beta$]{\includegraphics[width=1\columnwidth]{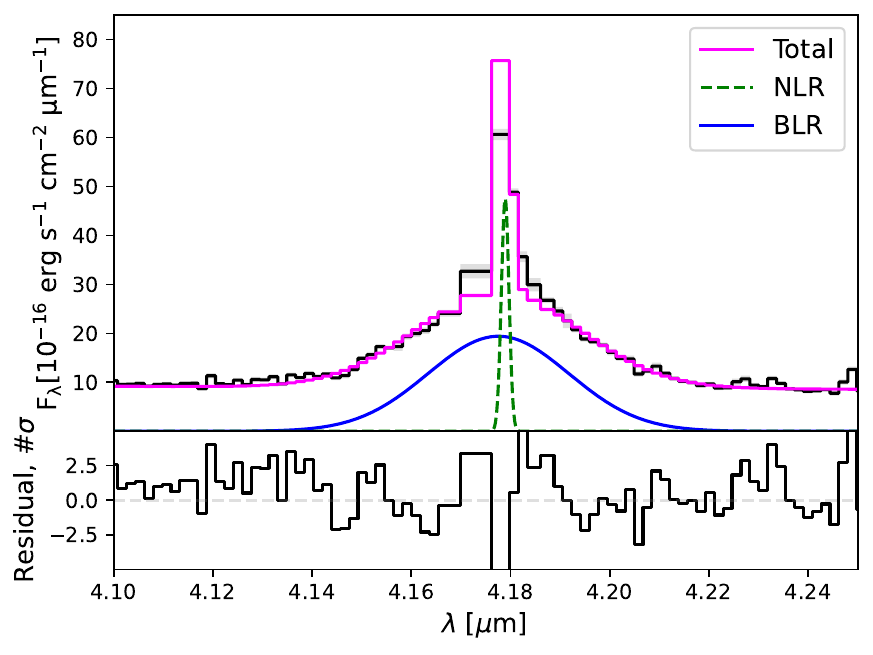}} \hfill
    \caption{R1000 spectrum expanded around some of the permitted lines of GN-28074 and showing the associated spectral fits. The broad line region (BLR) and the (extended) outflow components are plotted as solid blue and red lines respectively. The narrow components are shown with dashed green lines. The combined fit, including the absorption absorption components, is shown in magenta. \textbf{Top left: } \Hbs and \OIII$\lambda\lambda$5007,4959 complex, showcasing the deep absorption in \Hbs and broadening of \OIII$\lambda\lambda$5007,4959 lines due to the presence an outflow. \textbf{Top right: } \Has, \NII$\lambda\lambda$6548,6583 and \SII$\lambda\lambda$6716,6731 . Notable is the relative weakness of the \NII$\lambda\lambda$6549,6585 and \SII$\lambda\lambda$6716,6731 doublets. \textbf{Bottom left: } Combined \HeI$\lambda$10830 and Pa$\gamma$ line fit. The ionized outflow component is not required here by the data. \textbf{Bottom right: } Fit to the Pa$\beta$ line. Due to wavelength calibration issues, this line is slightly offset to the rest and thus requires separate kinematics to properly fit. The ionized outflow is absent just as in the Pa$\gamma$ + \HeI$\lambda$10830 fit.}
    \label{fig:spectrum_abs}
\end{figure*}

The spectrum in \autoref{fig:full_spec} shows clear broad components present in the bright permitted lines (hydrogen and helium lines), which lack kinematically comparable counterparts in the \OIIIs forbidden doublet. This unambiguously identifies their origin being the broad line region (BLR) of an active galactic nucleus (AGN). It should be noted that the profiles of \OIII$\lambda\lambda$4959,5007 and \SIII$\lambda$9531 lines do exhibit faint, broad wings, but these are likely associated with an ionized outflow as their kinematics differ strongly (much narrower) from the BLR and will be explored further in Appendix~\ref{appendix:outflow}.

In addition, the spectrum of GN-28074 clearly shows deep blueshifted (by a few 100 km s$^{-1}$) absorption features in the broad \Ha, \Hbs and \HeI$\lambda$10830 lines in \autoref{fig:spectrum_abs}. While the \HeI$\lambda$10830 is associated with a metastable level, hence often seen in absorption (provided that the conditions and column density are adequate), observing H$\alpha$ and H$\beta$ in absorption is much more difficult, as it requires significant population of the $n=2$ level, which has a very short lifetime ($\sim 10^{-8}$~s) for radiative transition. As we will show, this implies large densities and large column densities of the medium along the line of sight (LOS). 

The absorption features were modeled using a standard attenuation model:
\begin{equation}
\label{eq:absorption}
    f_{\lambda} = 1 - C_f + C_fe^{-\tau_{\lambda}},
\end{equation}
where $C_f$ is the covering factor of the absorber and $\tau_{\lambda}$ - its optical depth profile. The optical depth profiles were assumed to be Gaussian:
\begin{equation}
\label{eq:depth_prof}
    \tau_{\lambda} = \tau_0\exp{\left[-0.5\left(\frac{\lambda - \lambda_0e^{\Delta v/c}}{\sigma}\right)^{2}\right]},
\end{equation} 
with $\tau_0$ corresponding to the optical depth at the core of the line, $\Delta v$ - the velocity shift of the absorber and $\sigma$ - its velocity dispersion. The attenuation was applied to the BLR and continuum emission. Indeed, the absorption is so deep that it cannot be absorbing only the continuum (as subtracting the broad line emission would result in negative absorption), it must therefore be absorbing, at least partially, also the broad lines. This implies that the absorbing medium must be located within or just outside the BLR.

All emission lines in R1000 were modelled using Gaussian decomposition with separate Gaussians representing the narrow line, BLR and ionized outflow emission. Fitting was performed using a Bayesian method with linear uniform priors constraining the line widths to between 100 and 800~km~s$^{-1}$ for the narrow, 800 - 10,000~km~s$^{-1}$ for the broad and 600 - 2000~km~s$^{-1}$ for the ionized outflow components. The offset between the BLR and outflows was allowed to vary by up to $\pm$10,000~km~s$^{-1}$ with all BLR and ionized outflow components constrained to the same kinematics with the exception of \HeI$\lambda$10830 - due to its resonant nature its FWHM was treated separately. The widths of the broad Pa$\gamma$ and Pa$\beta$ lines were also left independent, because their intrinsic widths could be different from those of the Balmer series \citep[e.g.,][]{landt_pabawidth_2008}. In addition, the Pa$\beta$ line is affected by wavelength calibration issues causing a $\sim$1~px offset from the other lines, which forces us to leave all its kinematics, aside from the width of the narrow line, independent of the remaining lines. All linewidths were convolved with the line spread function (LSF) for point sources from \cite{DeGraaf2024} during the fitting procedure. Absorption in the Balmer lines was constrained to have the same kinematics, while the \HeI$\lambda$10830 absorption was fitted with independent $\Delta v$ and $\sigma$. The prior for $\Delta v$ was uniform, constrained to be between -1000 and 1000~km~s$^{-1}$, while $\sigma$ was allowed to vary between 0 and 400~km~s$^{-1}$. The covering factor was constrained to be the same for all absorption lines. At this stage (i.e. for the purpose of the line fitting) the continuum was approximated empirically as a fourth-order polynomial.

The posterior probability distribution of all model parameters was estimated using a Markov-Chain Monte-Carlo integrator \citep{emcee}; we initialised the chains in a small neighbourhood around the maximum-likelihood parameter values, estimated via least-squares minimisation. The moments of the posterior distribution for the neutral outflow model are presented in \autoref{tab:abs_par}. We note that \autoref{fig:spectrum_abs} displays significant residuals around the peaks of bright lines. The likely cause of this is the line intensity saturating the detector in some exposures resulting in the pipeline failing to derive valid errors for their peaks. A summary of all fitted emission line fluxes is provided in \autoref{tab:Line_summary} while the relevant line widths are summarized in \autoref{tab:kinematics}. The fit uncertainties for the narrow lines include a $\sim$20\% systematic uncertainty on the LSF, which is less significant for the well-resolved broad lines.
\begin{table*}
    \centering
    \renewcommand{\arraystretch}{1.5}
    \begin{tabular}{ccccccc}
    \hline
         Line & $C_f$ & $\tau_0$ & $\Delta v$~[km~s$^{-1}$] & $\sigma$~[km~s$^{-1}$] & $v_{out}$~[km~s$^{-1}$]& $\log{N}$~[cm$^{-2}$]\\
         \hline
         \HeI$\lambda$10830 & \multirow{3}{*}{$0.998_{-0.004}^{+0.001}$}& 2.17$^{+0.05}_{-0.05}$ & -506$_{-7}^{+7}$ & 304$_{-6}^{+6}$ & $1113_{-12}^{+12}$ &  $14.56^{+0.01}_{-0.01}$ \\
         \Ha & & 8.8$_{-1.3}^{+1.5}$ & \multirow{2}{*}{-351$_{-16}^{+14}$} & \multirow{2}{*}{126$_{-8}^{+6}$} &  \multirow{2}{*}{604$_{-10}^{+11}$} & $14.99^{+0.07}_{-0.07}$\\
         \Hb & & 1.02$_{-0.11}^{+0.12}$ & & & &$14.87^{+0.06}_{-0.06}$\\
         \hline
    \end{tabular} \quad
    \caption{A summary of absorption profile properties given by the fit. The first column gives the name of the fitted line, columns two and three - covering fraction and optical depth at the center of the absorption, columns four to five give the best-fit neutral outflow kinematics while the final column contains estimated column densities.}
    \label{tab:abs_par}
\end{table*}
\begin{table}
    \centering
    \renewcommand{\arraystretch}{1.3}
    \begin{tabular}{ccc}
    \hline
         Line & Type & Flux [$10^{-18}$~erg~s$^{-1}$~cm$^{-2}$]  \\
    \hline
        \OII$\lambda \lambda$3726,3729 & Narrow & $22.43_{-0.59}^{+0.62}$ \\
         H${\gamma}$ & Narrow & $10.48_{-0.29}^{+0.29}$ \\
         \OIII$\lambda$4363 & Narrow & $4.63_{-0.20}^{+0.18}$ \\
         \Hb & Narrow & $19.98_{-0.46}^{+0.45}$ \\
         \Hb & Broad & $32.1_{-1.1}^{+1.1}$ \\
         \Hb & Outflow & $7.58_{-0.93}^{+0.90}$ \\
         \OIII$\lambda$5007 & Narrow & $146.8_{-1.0}^{+1.0}$ \\
         \OIII$\lambda$5007 & Outflow& $14.51_{-0.61}^{+0.58}$ \\
         \OI$\lambda$6300 & Narrow& $2.38_{-0.14}^{+0.14}$ \\
         \Ha & Narrow & $82.64_{-0.27}^{+0.30}$ \\
         \Ha & Broad & $297.2_{-2.6}^{+2.7}$ \\
         \Ha & Outflow & $144.8_{-2.5}^{+2.3}$ \\
         \NII$\lambda$6583 & Narrow & $0.96_{-0.35}^{+0.34}$ \\
         \SII$\lambda$6716 & Narrow & $1.10_{-0.10}^{+0.10}$ \\
         \SII$\lambda$6731 & Narrow & $0.83_{-0.08}^{+0.08}$ \\
         \OII$\lambda \lambda$7320,7331 & Narrow & $1.50_{-0.25}^{+0.26}$ \\
         OI$\lambda$8446 & Narrow & $2.33_{-0.20}^{+0.20}$ \\
         \SIII$\lambda$9531 & Narrow & $4.39_{-0.22}^{+0.21}$ \\
         \SIII$\lambda$9531 & Outflow &  $22.3_{-1.2}^{+1.1}$\\
         \HeI$\lambda$10830& Narrow & $22.71_{-0.52}^{+0.51}$ \\
         \HeI$\lambda$10830& Broad & $59.4_{-1.3}^{+1.2}$ \\
         Pa$\gamma$ & Narrow & $5.11_{-0.23}^{+0.21}$  \\
         Pa$\gamma$ & Broad & $54.6_{-1.1}^{+1.3}$  \\
         OI$\lambda$11287 & Narrow & $1.68_{-0.20}^{+0.20}$ \\
         Pa$\beta$ & Narrow & $16.06_{-0.42}^{+0.44}$  \\
         Pa$\beta$ & Broad & $104.0_{-1.6}^{+1.6}$  \\
    \hline
    \end{tabular}
    \caption{Table summarizing all fitted emission line fluxes as measured - without correcting for dust attenuation. The first column indicates the name of a given line, the second - its type (Narrow, Broad or Outflow). The final column gives the measured flux in units of $10^{-18}$~erg~s$^{-1}$~cm$^{-2}$.}
    \label{tab:Line_summary}
\end{table}

\begin{table}
    \centering
    \renewcommand{\arraystretch}{1.3}
    \begin{tabular}{cc}
    \hline
         Width~[km~s$^{-1}$] & Value  \\
    \hline
         FWHM$_{\rm br}$ & $3610_{-22}^{+21}$ \\
         FWHM$_{\rm nr}$ & $151_{-36}^{+36}$ \\
         FWHM$_{\rm of}$ & $1311_{-19}^{+18}$ \\
         FWHM$_{\rm Pa_{\gamma}}$ & $2300_{-50}^{+53}$ \\
         FWHM$_{\rm Pa_{\beta}}$ & $2820_{-31}^{+31}$ \\
         FWHM$_{\rm HeI}$ & $1860_{-28}^{+27}$ \\
         \hline
    \end{tabular}
    \caption{Summary of widths for all fitted emission lines. The first three rows list the common widths fitted to the broad, narrow and outflow lines. The remaining rows give FWHM for the lines for which it was left independent.}
    \label{tab:kinematics}
\end{table}

\section{Continuum fitting and mid-IR excess}

In the NIRCam images GN-28074 appears as a point source with a spatially extended component to the south-east (inset single-band cutouts in \autoref{fig:full_spec}). This extension is clearly seen in all available filters from the SW NIRCam module (F090W--F210M), but we do not rule out its presence at redder wavelengths (an in-depth morphological analysis is beyond the scope of this article).
Shifting focus on the inset RGB image, we infer that this south-east extension is very likely to be at the same redshift as GN-28074. The extension has a strong line excess in F210M relative to F182M (and F210W). Notice that this excess feature (pink in the inset RGB) is stronger than what is seen at other position angles around the central source, which indicates it is an SED feature of the south-east extension, not of the point source itself. The most parsimonious explanation is that the line excess in F210M is tracing H$\alpha$ in the south-east region, which, therefore, is at the same redshift as the point source.
The pink colour in the RGB image also indicates that, compared to the point source, this extended component is stronger in F210M than in the rest-UV and rest-optical continuum. This however could result from at least two different scenarios: an SED dominated by star formation, or an SED dominated by nebular emission, such as the ionisation cone of an AGN \citep[e.g.,][]{Tacchella2024}.
Interestingly, F150W seems to be more extended than F182M. Given that the NIRCam PSF is broader in F182M than F150W, this line of evidence may imply the presence of a diffuse region of \OIII\ emission.

In a recent article, \citet{Wang2024} report the discovery of a broad-line AGN with weak or even no dust emission, confirmed by MIRI. In addition, \citet{Matthee2024} find a tentative detection of broad H$\alpha$ (4$\sigma$) that was confirmed as MIR-weak by \cite{Williams2024}, contrary to expectations that similar sources should exhibit strongly rising MIR continuum from AGN dust heating \citep[e.g.][]{Labbe2023}. While a weak MIRI SED appears prevalent among photometrically identified samples \citep[e.g.][]{Williams2024, Akins2024, Perez-Gonzalez2024}, unfortunately, we still lack extensive MIRI coverage of spectroscopically confirmed type 1 AGN, therefore we cannot yet verify if the lack of MIR emission is a peculiarity of one--two sources, or if it is a more common feature of low-luminosity AGN, perhaps due to a lack of a dusty torus, or due to colder dust temperatures than in previously studied AGN. To investigate if GN-28074 also lacks MIR emission, we utilize Spitzer IRAC imaging across all four channels, covering 3.6 to 8~$\mu$m, together with MIPS 24~$\mu$m band. Note that MIPS may suffer from contamination due to a nearby source, and that we do not include \textit{Herschel} non detections, because their high upper limits are not constraining. The rest-frame UV and optical range was covered using JWST and HST photometry, as discussed in section 2. Our object has S/N > 10 across all bands considered allowing for robust SED fitting of the rest-frame UV to near infrared continuum. This also marks GN-28074 as one of the few JWST AGN to have robust MIR constraints.

We carry out the fits using the \textsc{CIGALE} code \citep{cigale}. The AGN component of the templates was modeled using \textsc{SKIRTOR} continuum emission models \citep{Skirtor_mods} with AGN fraction allowed values of 0 to 0.99, the polar dust extinction of the continuum was modeled with the SMC extinction law, standard for dust reddened AGN \citep{Reichard2003, Richards2003}, and allowed E(B - V) values ranging from 0.03 to 4. Stellar populations were modeled using BC03 templates with \cite{Salpeter-IMF} IMF. Stellar dust attenuation was modeled using the Calzetti attenuation law with E(B-V) allowed to vary between 0 and 4. Dust emission was modeled with \cite{Dale2014} models with emission slope $\alpha = 2$. Gas and stellar metallicities were set to either solar or ~10\% solar, however, the exact metallicity has little impact on the shape of the model SEDs. Parameters not mentioned were kept to \textsc{CIGALE} defaults.

\begin{figure*}
    \centering
    \includegraphics[width=\textwidth]{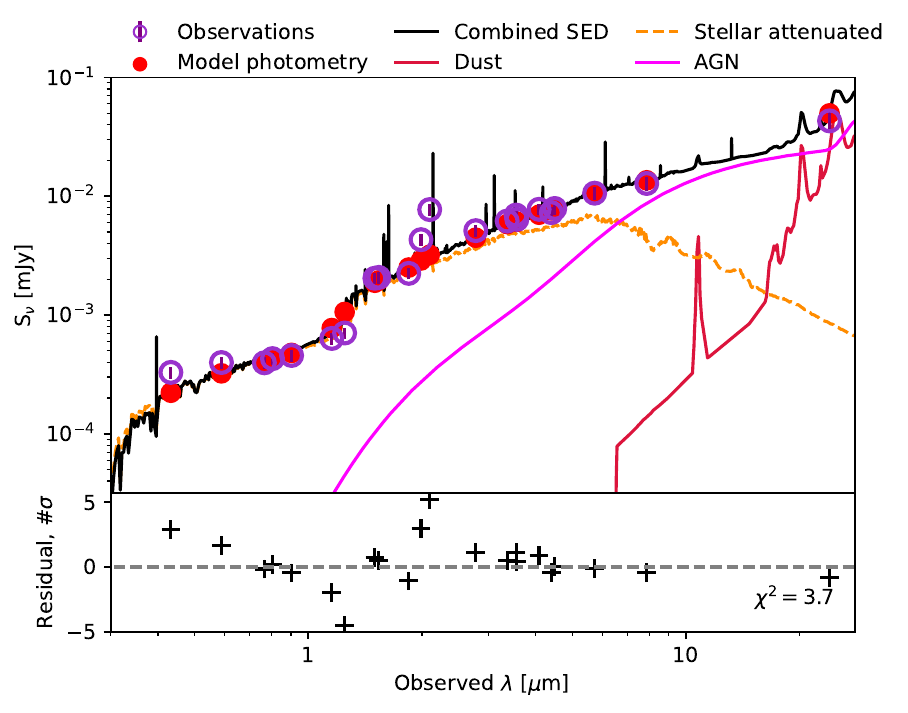}
    \caption{The best-fitting \textsc{CIGALE} model. It can be seen that the mid infrared emission is well explained by hot dust emission from an attenuated accretion disk emission (magenta) with comparatively little contribution from stellar dust (in red). The excess flux around 2.0$\mu$m that is not well reproduced can be attributed to the strong \Has emission.}
    \label{fig:AGN_SED}
\end{figure*}

The resulting fit is shown in \autoref{fig:AGN_SED} and shows that the best-fitting model prefers the presence of a strongly attenuated accretion disk component with $\rm{E(B-V)} = 1.04 \pm 0.24$, corresponding to an $A_V = 3.02 \pm 0.70$, with accompanying infrared emission from hot dust forming the continuum emission displayed there in magenta. The stellar continuum component is attenuated considerably less, $A_V = 0.68 \pm 0.01$.

\begin{figure*}
    \centering
    \includegraphics[width=\textwidth]{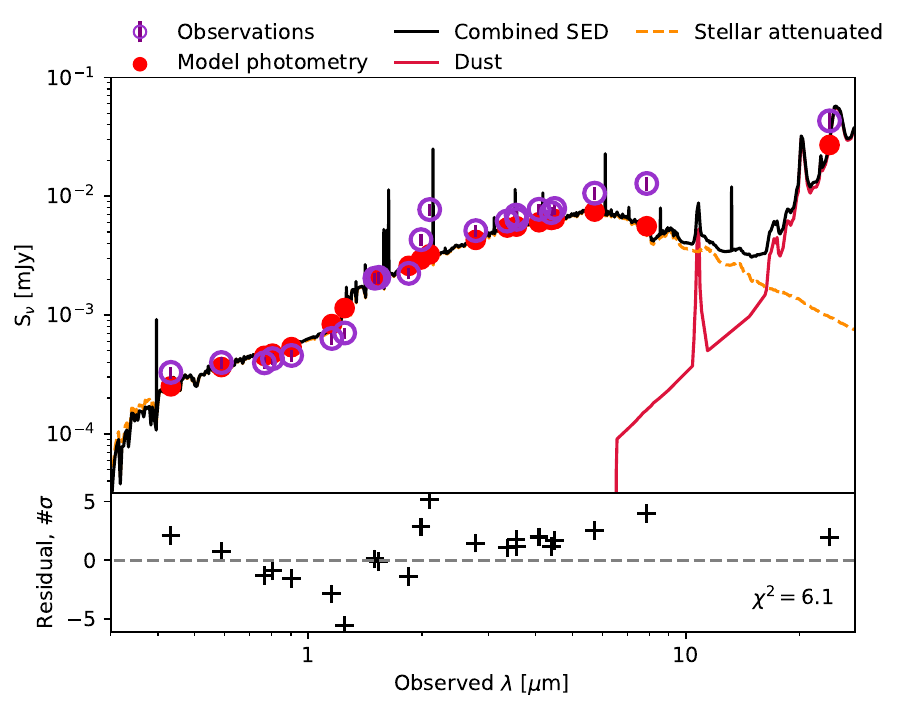}
    \caption{Same as \autoref{fig:AGN_SED}, except for models without AGN emission. It can readily be seen that stellar-only templates leave significant residuals from 5$\mu$m onwards resulting in a poorer reduced $\chi^2$ value.}
    \label{fig:stellar_sed}
\end{figure*}

To test if the mid infrared photometry can be explained with stellar emission or associated dust reprocessing alone we refit the data by forcing the AGN fraction to 0. The resulting best fit without AGN is shown in \autoref{fig:stellar_sed}. As can be seen, the stellar-only models yield a considerably worse overall fit and can not explain the infrared photometry, necessitating the presence of a hot dust component, which can be readily identified as a dusty torus or hot polar dust. No matter the exact interpretation, our result indicates the presence of significant amounts of hot dust surrounding the core of GN-28074.

A notable feature, common to both stellar and AGN+stellar fits, is that both of them attempt to fit the discontinuity in flux between 1 and 2 $\mu$m with a Balmer break. However, both leave a significant residual in that location (\autoref{fig:AGN_SED}, \autoref{fig:stellar_sed}), thus a Balmer break fit does not explain the emission pattern well. If the feature is indeed a Balmer break, indicative of a more evolved stellar population, the stellar mass estimate gives $\log{\frac{M_*}{M_{\odot}}} = 10.78 \pm 0.05$. For comparison, we use the fits to the narrow lines to estimate an upper limit on the dynamical mass of the object using the virial estimator (calibrated on stellar kinematics) from \citet{cappellari+2006}:
\begin{equation}
    M_\mathrm{dyn} < \frac{5\sigma^2R_e}{G},
\end{equation}
where $\sigma$ is the gas velocity dispersion and $R_e$ the circularised half-light radius. The fits to the narrow lines give a FWHM = $150_{-36}^{+36}$~km~s$^{-1}$, corresponding to $\sigma \approx 64$~km~s$^{-1}$. The value for $R_e$ is trickier to estimate as the object is point-source dominated, thus we adopt 0.06~arcsec as an upper limit, using the value measured by \cite{vanderWel2014} from HST F160W imaging. This value is the semi-major axis, which translates into a circularised $R_\mathrm{e}=0.035$~arcsec after taking into account the measured shape \citep[axis ratio $q=0.65$;][]{vanderWel2014}. The resulting mass is $\log{\frac{M_\mathrm{dyn}}{M_{\odot}}} \lesssim 9.1$, lower than the `maximal' stellar mass inferred interpreting the spectral break as a Balmer break.
The strong mismatch with $M_*$ does not depend on the specific virial calibrator we use; alternative formulae give $\log{\frac{M_\mathrm{dyn}}{M_{\odot}}} = 8.9\text{--}9.5$ \citep{wisnioski+2018,vanderwel+2022}. This inconsistency could be due to geometric effects, if the narrow-line gas kinematics are dominated by a disc seen close to face on. However, \citet{Wang2024_z7} reports three more examples of broad-line AGN with spectral breaks around rest-frame 3,600--4,000~\AA, and where the virial mass is lower than the maximal stellar mass. Therefore, it is unlikely that the observed drop in the spectrum can be explained entirely by a Balmer break. We further note that in the R1000 spectrum (\autoref{fig:full_spec}) the break appears less steep and at redder wavelengths than the typical Balmer break. However, interpreting the break as a stellar 4000-\AA\ break would further increase the stellar mass and, consequently, the tension with the dynamical-mass constraint.

We do not discuss further the nature of the drop between 3,000 and 4,000 \AA\ rest frame, as we defer further analysis to another paper in which other similar sources are also analysed.

\section{Derived quantities}
\label{sec:analysis}
\subsection{Black hole properties}
\label{subsec:BH_prop}

We utilize our fits to the broad line region (BLR) emission to constrain the black hole mass using single-epoch virial relations. Recent results by \cite{Gravity24} measured the BLR size via interferometric data, and have put into question the usage of BH virial relations using transitions such as CIV \citep{Netzer2007, Negrete2012}; however, their finding that the discrepancy is reduced to only a factor 2.5 when using the \Has line is reassuring, as this is  within the calibration uncertainties. Furthermore, the deviation is associated with super-Eddington accretion influencing the size of the BLR, while the BH of 28074 is likely in a sub-Eddington regime.

Our fits to the BLR yield $\rm FWHM_{H\alpha} = 3610_{-22}^{+21}$~km~s$^{-1}$ along with $L_{\rm H\alpha} = 1.21_{-0.01}^{+0.01} \times 10^{43}$~erg~s$^{-1}$. Converting this to black hole mass via the virial relation \citep{VolonteriBHmass}:
\begin{multline}
\label{eq:virial_mass}
    \log{\frac{M_{\rm BH}}{M_{\odot}}} = \\
    6.60 + 0.47\log{\left(\frac{L_{H\alpha}}{10^{42}\ {\rm erg\,s^{-1}}}\right)+2.06\log\left(\frac{{\rm FWHM}_{H\alpha}}{1000 {\rm km\,s^{-1}}}\right)} 
\end{multline}
yields $\log{\frac{M_{\rm BH}}{M_{\odot}}} = 8.26 \pm 0.30$, the uncertainty on this value is completely dominated by the intrinsic scatter on \autoref{eq:virial_mass}. The scaling relation between the luminosity of the broad \Has component and bolometric luminosity calibrated by \cite{SternLbol} implies $L_{\rm bol} = 1.58_{-0.01}^{+0.01}\times 10^{45}$~erg~s$^{-1}$, with a 0.3~dex intrinsic scatter, and an Eddington ratio of $\sim$7\%. 

Estimating the extinction correction to the BLR is not trivial even when a broad component in both \Hbs and \Has lines is seen, as the intrinsic Balmer decrement can be significantly steeper than the case B scenario and can potentially reach values up to ten \citep{Ilic2012}. In the case of our object, we find \Has to \Hbs ratios to be $9.25_{-0.31}^{+0.33}$ and $4.14_{-0.14}^{+0.16}$ for the broad and narrow lines respectively. Assuming a standard case B decrement of 2.85 for the narrow lines along with the SMC extinction curve \citep{smc_gordon}, which might be representative in galaxies at $z\sim 2$ \citep[e.g.,][]{reddy2015}, we obtain $A_V = 1.13_{-0.10}^{+0.11}$. Adopting this value for the dust correction increases the measured $L_{bol}$ by $\sim$ 0.5~dex and results in a corrected BH mass of $\log{\frac{M_{\rm BH}}{M_{\odot}}} = 8.47 \pm 0.30$ and an Eddington ratio of $0.12$. If we instead take the broad lines Balmer ratio and conservatively assume an intrinsic case B ratio, we would get an upper limit on the extinction of $A_V<2.3$, which would imply a maximum correction on the bolometric luminosity of 0.9~dex, a maximum black hole mass of $\log{\frac{M_{\rm BH}}{M_{\odot}}} < 8.67$, and a maximum Eddington ratio of 0.2.
One should also take into account the additional uncertainty on both black hole mass and Eddington ratio of 0.3 -- 0.5~dex of the virial relations. Observed properties of the BLR are summarized in \autoref{tab:BLR}.

\begin{table}
    \centering
    \renewcommand{\arraystretch}{1.3}
    \begin{tabular}{ccc}
    \hline
        Quantity & Best estimate & Upper limit  \\
    \hline
        FWHM$_{\rm H\alpha}$~[km~s$^{-1}$]& $3610_{-22}^{+21}$ & --\\
         $L_{\rm H\alpha}$~[erg~s$^{-1}$] & $3.44_{-0.31}^{+0.39}\times10^{43}$ & $<2.1\times 10^{44}$ \\
         $L_{\rm H\beta}$~[erg~s$^{-1}$]& $5.36_{-0.65}^{+0.87}\times10^{42}$ & $<7.3\times 10^{43}$\\
         $L_{\rm bol}$~[erg~s$^{-1}$]& $4.47_{-0.41}^{+0.50}\times10^{45}$ & $<2.7\times 10^{46}$\\
         $\log{M_{\rm BH}}$~[M$_{\odot}$]& $8.47 \pm 0.30$ & $<8.67$\\
         $\lambda_{\rm Edd}$ & $0.120_{-0.006}^{+0.007}$ & $< 0.2$ \\
         \hline
    \end{tabular}
    \caption{A summary of the observed properties of the BLR and the inferred mass of the supermassive black hole and its accretion rate. The first column denotes the name of the quantity, the second - its value. Listed luminosities are corrected for dust. The final column provides upper limits based on the conservative A$_V$ estimation based on the observed broad Balmer line decrement.}
    \label{tab:BLR}
\end{table}

\subsection{Nuclear neutral absorption}

\subsubsection{Modelling and evidence for very high gas densities}

From \autoref{fig:spectrum_abs}, it is clear that strong blueshifted absorption features are present on top of the broad components of \Hb, \Ha, and \HeI$\lambda 10830$. 
Before JWST,
strong blueshifted absorptions in hydrogen and/or helium lines have previously been observed in the spectra of a few AGN
\citep[e.g.,][]{hutchings2002,aoki2006,aoki2010,hall2007,ji2012,ji2013,wangxu2015,Zhang2015,Shi2016,williams2017,Schulze2018,hamann2019}, including the prototypical local Seyfert 1 NGC4151 \citep{hutchings2002}. However, these were extremely rare cases, less than 0.1\% of the quasar/AGN population.
Yet, the JWST spectra are revealing an increasing number of these Balmer line absorption features in Type 1 AGN, possibly making about 10\% of the AGN population discovered by JWST
\citep{Matthee2024,Wang2024,Wang2024_z7,Kocevski2024}. Therefore, such absorption features are probably the signature of properties characterizing the new population of AGN discovered by JWST.
The prominence of these features in GN-28074, along with the wealth of additional information, makes this object the perfect case study to explore these properties.

As already discussed, absorption of \Ha\ and \Hb, and, to some extent, \HeI, requires peculiar conditions of the gas along the LOS.
Specifically, these need substantial amount of (mostly) neutral gas along the LOS that has significant amounts of hydrogen
and helium pumped to excited states, which are the $n=2$ level for hydrogen and the metastable triplet state $2^3S$ for helium. Due to the short lifetime of H($n=2$) compared to that of He($2^3S$), it is much more difficult to have significant absorption in hydrogen lines without the presence of high-density gas, as we demonstrate in the following.

We estimate the absorbing column densities {of excited hydrogen and helium} by integrating \autoref{eq:depth_prof} with the best-fit parameters using the apparent optical depth method \citep{Wang2024, SavageSembach1991}. The column density is then given by:

\begin{equation}
    N = \frac{m_ec}{\uppi e^2 f_0\lambda_0}\tau,
\end{equation}
where $\tau$ is the integrated {line} optical depth and $f_0\lambda_0$ is the product of the rest-frame wavelength and the oscillator strength of a given transition. Using this framework we find that the absorption in \Has and \Hbs give excited ($n=2$) hydrogen column densities of $\log{N_{\rm H(n=2)}} = 14.99^{+0.07}_{-0.07}$~cm$^{-2}$ and $14.87^{+0.06}_{-0.06}$~cm$^{-2}$ respectively, which are fully consistent between one another. The absorption profile of \HeI$\lambda$10830 yields $\log{N_{\rm HeI^*}} = 14.56^{+0.01}_{-0.01}$~cm$^{-2}$.
Given the relatively short lifetime of the excited hydrogen at the $n=2$ level, its large column density indicates a high total column density of hydrogen \citep{agn3}.

The absorption line kinematics of the fit gives $\sigma = 126_{-8}^{+6}$~km~s$^{-1}$ and $\Delta v = -351_{-16}^{+14}$~km~s$^{-1}$ for the Balmer lines and $\sigma = 304_{-6}^{+6}$~km~s$^{-1}$ and $\Delta v = -506_{-7}^{+7}$~km~s$^{-1}$ for the helium absorption lines. Estimating maximum outflow velocity as $v_{\rm out} = |\Delta v| + 2\sigma$ gives $v_{\rm out} = 604_{-10}^{+11}$~km~s$^{-1}$ and $v_{\rm out} = 1113_{-12}^{+12}$~km~s$^{-1}$ for the hydrogen and helium outflows respectively.
We summarize our derived neutral gas outflow properties in \autoref{tab:abs_par}.

While the outflow velocities inferred from the hydrogen and helium absorptions are significantly different, we note that the resonant nature of \HeI$\lambda 10830$ could well lead to different kinematics and it is thus not a direct reflection of the outflow velocity. 
One might wonder whether the absorbing gas could have two components: one faster outflowing and low-density component primarily responsible for the absorption of \HeI$\lambda 10830$, and the other slower outflowing and high-density component primarily responsible for the absorption of hydrogen Balmer lines.
Regardless, we emphasize that a dense neutral outflow must exist, as we explain later in this section.

\begin{figure}
    \centering
    \includegraphics[width=\columnwidth]{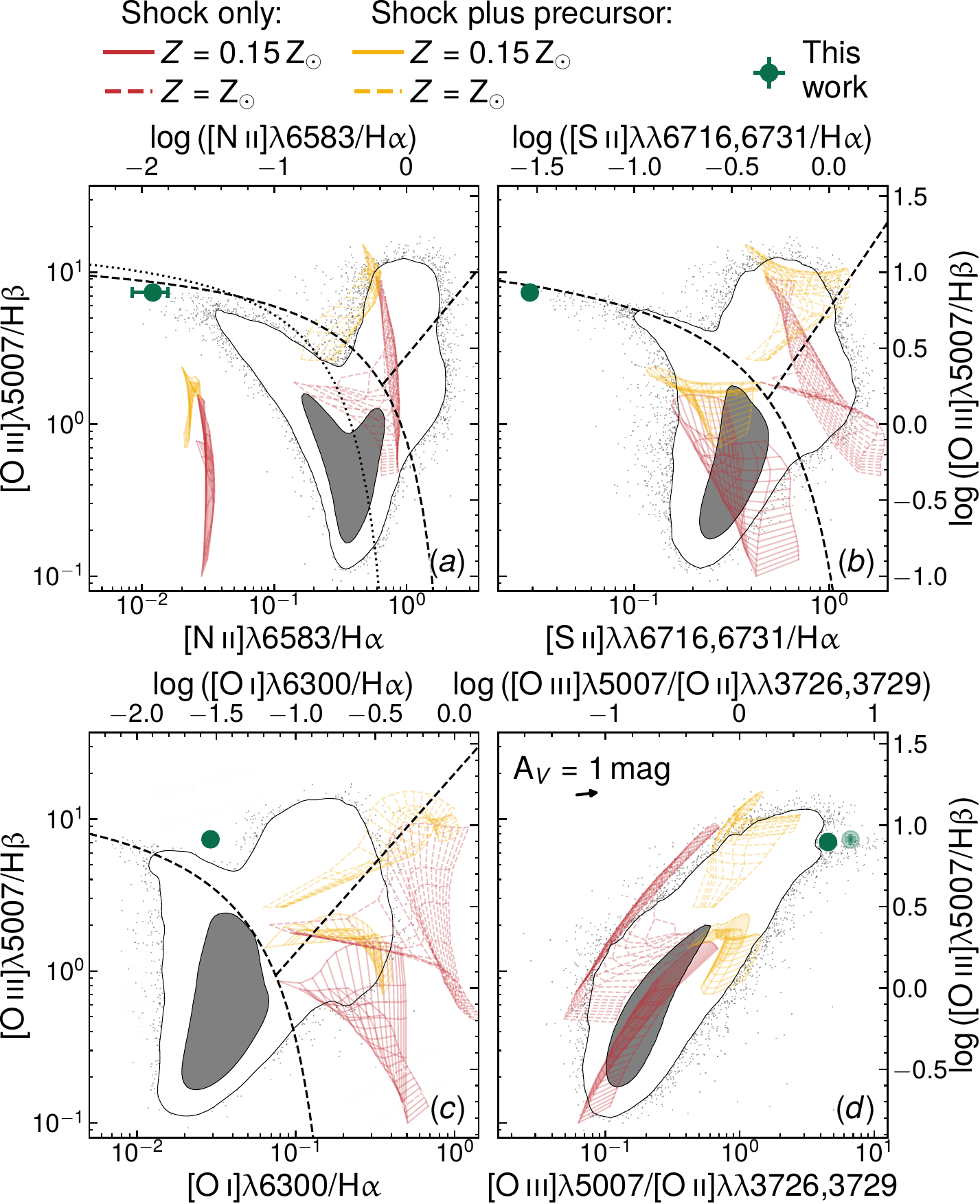}
    \caption{Narrow emission-line diagnostic diagrams, showing our target (green circles with errorbars; the measurement uncertainties are often smaller than the size of the marker), and the distribution of galaxies at $z=0.1$ from SDSS Data Release 7 (contour lines and scatter points). The demarcation lines in panels a--c are from \protect\cite{kewley+2001, kauffmann+2003c, kewley+2006, schawinski+2007}. Our target lies in a scarcely populated region of the diagram, formally consistent with star-formation photoionisation (panels~a and~b) or AGN (panel~c). Shock-driven emission in the NLR is also unlikely, because shock models (red and yellow lines) do not cover the location of our target in these diagnostic plots (see text for a description).}
    \label{fig:shock_bpt}
\end{figure}

\begin{figure*}
    \centering\includegraphics[width=.95\columnwidth]{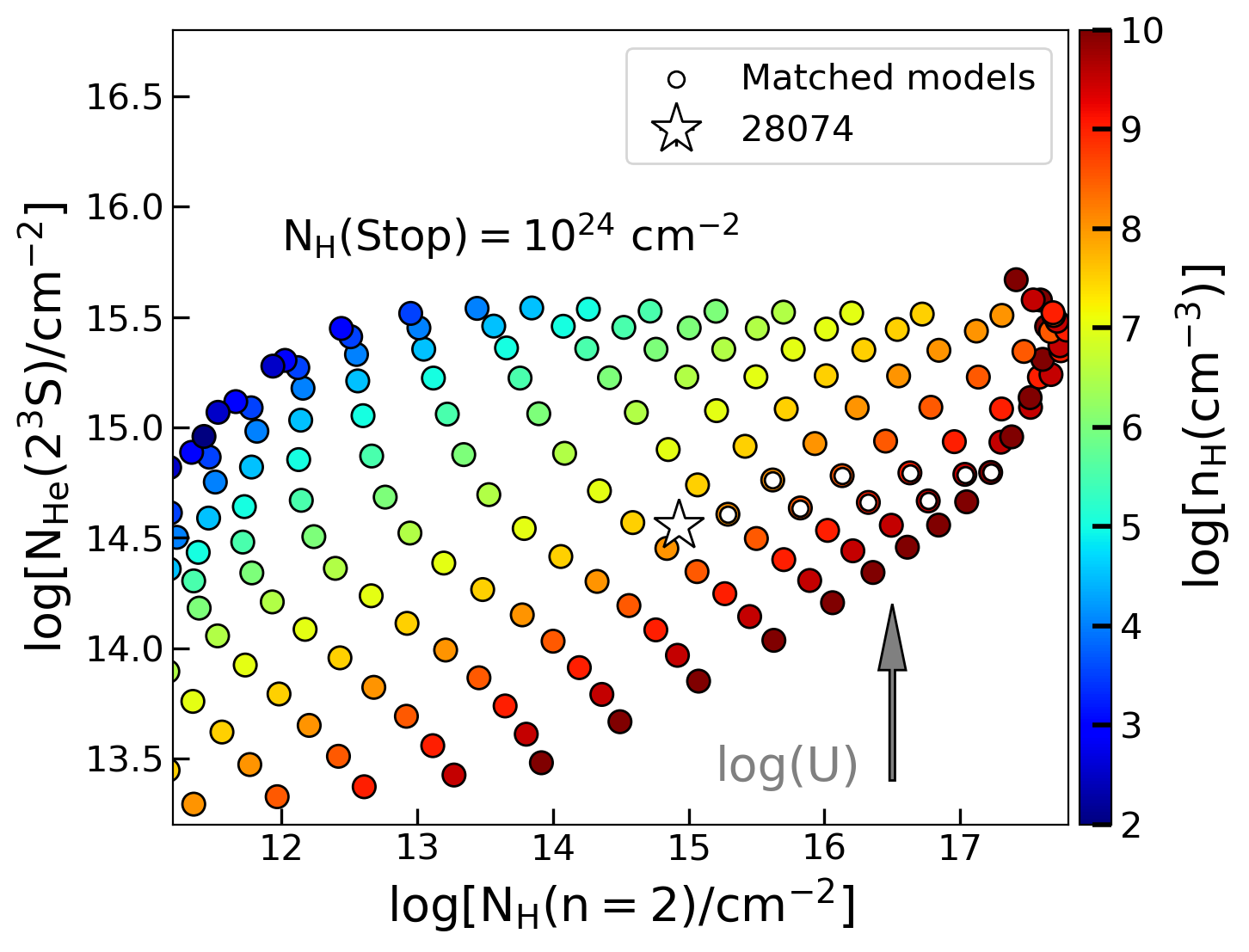}
    \centering\includegraphics[width=.97\columnwidth]{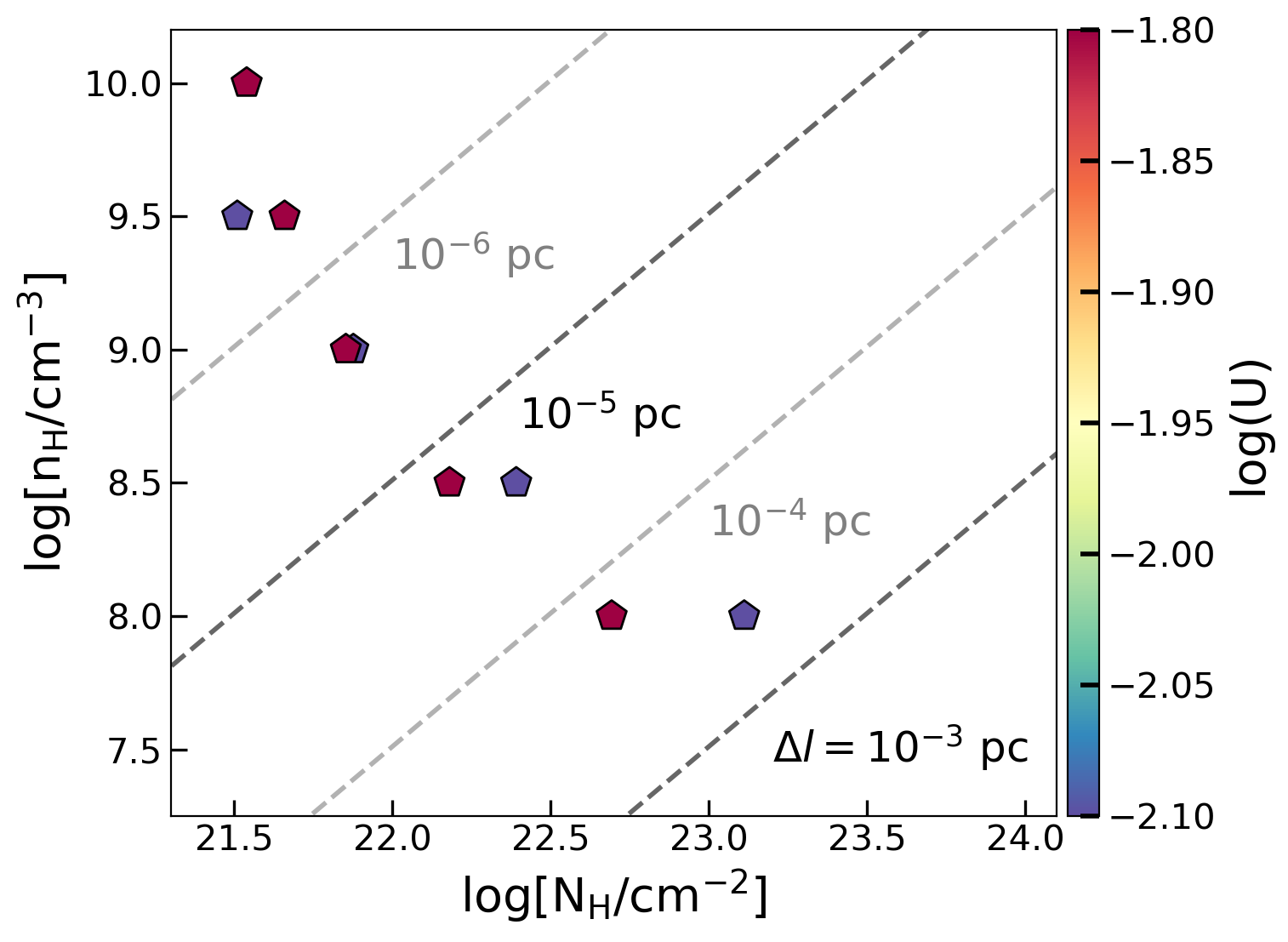}
    \caption{Photoionization models of pumped absorbing gas that has high column densities of hydrogen at the $n=2$ state and helium at the first excited state. 
    \textit{Left:} $N_{\rm H(n=2)}$ and $N_{\rm He(2^3S)}$ at a common depth of $N_{\rm H}{\rm (Stop)} = 10^{24}~{\rm cm}^{-2}$ for models spanning a range of hydrogen densities and ionization parameters.
    All models are plotted as circles color-coded by their hydrogen densities.
    The grey arrow indicates the increasing direction of the ionization parameter in this diagram.
    The star symbol represents our derived column densities for GN-28074.
    The open circles represent models that are capable of reproducing our derived column densities for GN-28074 at certain depths.
    The plausible models have high hydrogen densities and cover a narrow range in the ionization parameter.
    \textit{Right:} hydrogen densities and column densities of plausible models colored coded by their ionization parameters.
    The dashed lines correspond to constant effective thickness of the absorbing gas.
    All the plausible models show small effective thickness of $\Delta l < 10^{-3}$ pc.
    }
    \label{fig:model_sol}
\end{figure*}

To infer the physical conditions within the neutral outflow, we compute a series of simulations using the photoionization code \textsc{Cloudy} \citep[c17.03;][]{cloudy} and compare them with observations. Specifically, we search for models with column densities in the $n=2$ level of the hydrogen atom and the first excited state $2^3S$ of the helium atom that match our derivations.

To start with, we need to assume what is the energy source that pumps the hydrogen and helium in the absorber to their excited states.
A natural candidate is the energetic photons from the AGN continuum emission in GN-28074.
In principle, powerful shocks, as seen in some radio AGN, can also create excited hydrogen \citep[e.g.,][]{best2000}.
However, a shock in the NLR is heavily disfavoured by the BPT/VO diagrams \citep{bpt,vo} as shown in \autoref{fig:shock_bpt}: our target (green point with errorbars) is clearly inconsistent with a broad range of shock models indicated with yellow and red grid lines\footnote{The models are from \textsc{mappings\;v}, \citealp{sutherland+dopita2017}; they assume the solar abundances of \citealp{gutkin+2016} and were obtained from the Mexican Million Models Database, 3MdB, \citealp{alarie+morisset2019}; the red and yellow grids display predictions for shock-only and for shock-and-precursor models, \citealp{allen+2008}}.

On the other hand, narrow line diagnostics can not be used to rule out a shock front in the BLR.
However, shocks produced by radio jets (primary candidates for producing strong BLR shocks) are strongly disfavoured by the radio non detection, as discussed later on in Section~\ref{subsec:radio_nondetection}.


We use a standard AGN spectral energy distribution (SED) in \textsc{Cloudy} as the source to pump hydrogen and helium to their excited states. The SED has a parametric form given by
\begin{equation}
    F_\nu = \nu ^{\alpha _{\rm uv}}{\rm exp}(-h\nu /kT_{BB}){\rm exp}(0.01~{\rm Ryd}/h\nu) + a\nu ^{\alpha _{\rm x}},
\end{equation}
where we set the big blue bump temperature to $T_{BB} = 10^{6}$ K, the UV-to-X-ray slope to $\alpha _{\rm ox} = -1.4$ (which is achieved by adjusting $a$ by the code), the UV slope to $\alpha _{\rm uv}=-0.5$, and the X-ray slope to $\alpha _{\rm x} = -1.0$
\footnote{While we cannot well constrain the actual spectral slope of GN-28074 due to the potentially significant dust attenuation, we have tried varying the optical-to-X-ray slope as well as the temperature of the accretion disk over a reasonable range, which, however, does not change our conclusions.}.
We compute a total of 272 models spanning a range of hydrogen densities of $10^2~{\rm cm^{-3}}\leq n_{\rm H} \leq 10^{10}~\rm cm^{-3}$, and a range of ionization parameters of $10^{-4.5} \leq U \leq 10^{0}$. The dimensionless ionization parameter, $U$, is defined as $U\equiv \frac{\Phi _0}{n_{\rm H}c}$, where $\Phi _0$ is the number flux of the hydrogen ionizing photons (impinging on the inner face of the cloud) and $c$ is the speed of light.
For all of the models, we adopt a metallicity of $\rm 12+log(O/H) = 7.85$ (which is derived using narrow line fluxes of GN-28074, as described in Appendix~\ref{appendix:met_ion_outflow}), and assuming the solar abundance pattern follows the values in \citet{solar}.
While the metallicity in the absorber might be different from that in the forbidden-lines emitting region, overall the metallicity has negligible impact on our conclusions (as we are focusing on hydrogen and helium absorption).
We do not include dust grains in our models.
While dust can also absorb ionizing photons, at low metallicities, the amount of dust is significantly reduced in the ISM and its impact on the ionization structure of the gas becomes negligible.
Furthermore, as we show later, the absorber has a density similar to the gas in the BLR, which is usually believed to be dustless (because it is inside the sublimation radius).

We have included the effect of microturbulence in our simulations and varied the turbulent velocity over a range of $\rm 100~km\, \mathrm{s}^{-1} \leq v_{turb} \leq 300~km\, \mathrm{s}^{-1}$ to match the velocity dispersions we fit for the hydrogen and helium absorptions. The microturbulence broadens the line profile and increases the chance of absorption. However, within the range of turbulent velocity we choose, the exact value of the turbulent velocity has little impact on our final results.

Finally, all models are truncated at a total hydrogen column density of $N_{\rm H} = 10^{24}~{\rm cm^{-2}}$. We search within each model whether there is a specific depth at which the cumulative column densities of the excited hydrogen and helium match observations within $2\sigma$ ($\sim 0.14$~dex).

{The left panel of \autoref{fig:model_sol} shows the total
column densities of excited hydrogen (at $n=2$) and helium (at $2^3S$) in our models at a common depth of $N_{\rm H} = 10^{24}~{\rm cm^{-2}}$. {Since this is the common stopping criterion of our models, the column densities shown are the highest values achievable by the models.}
The models are color-coded by their hydrogen volume densities and the direction of increasing ionization parameter is indicated by the arrow.
The models that can reproduce the column densities we derive for GN-28074 (open star) within certain depths, quantified by column densities reached ($\lesssim N_{\rm H} = 10^{24}~{\rm cm^{-2}}$), are highlighted by open circles (indeed the different combinations of $N_{\rm H(n=2)}$ and $N_{\rm He}(2^3S)$ are degenerate in reproducing the observed absorptions equally well).
Clearly, the high $N_{\rm H(n=2)}$ we derive for GN-28074 strongly constrains the hydrogen density of the absorber to high values.
This is because higher densities allow $N_{\rm H(n=2)}$ to build up at smaller depths.
In addition, high densities boost collisional excitation of hydrogen, which helps to achieve high $N_{\rm H(n=2)}$.
Meanwhile, our derived $N_{\rm He(2^3S)}$ further narrows down the plausible range of densities and strongly constrains the ionization parameter.
}

With these results, we go back to consider the possibility that the Balmer absorption and the \HeI\ absorption are produced by different absorbing clouds.
From \autoref{fig:model_sol}, one can see the $N_{\rm H(n=2)}$ of GN-28074 alone would indicate a high $N_{\rm He(2^3S)}$ in the cloud that produces the Balmer absorption, unless the density is even higher.
Therefore, the most likely picture would be either a single absorber making both absorptions, or two separate absorbers with the Balmer-line absorber having a higher density.
Either way, the hydrogen density is constrained to be $n_{\rm H} \gtrsim 10^{8}~{\rm cm^{-3}}$, indicative of a BLR origin.
In what follows, we set our discussion in the context of a single absorber for simplicity, and we note that the more complex two-cloud possibility would not change our main conclusions.


In the right panel of \autoref{fig:model_sol}, we show the densities and column densities of the plausible absorbing models for GN-28074. We note that the column density in this panel is \textit{not} the total column density of the models (which has been fixed to $N_{\rm H} = 10^{24}~{\rm cm^{-2}}$), but the cumulative hydrogen column densities in certain depths of the models where both column densities of He ($2^3S$) and H ($n=2$) match the values of GN-28074. 
These are the minimum values of the actual column density of the cloud in order to match the observation, as the cloud can have an additional column of gas not playing a role in the absorption seen in the Balmer and \HeI\ lines (because the $n=2$ and $2^3S$ levels are not properly excited in these regions).
The models show an anti-correlation between the density and column density, which is expected since higher densities help to populate hydrogen to the $n=2$ level through collisional excitation.
Overall, the plausible hydrogen (minimum) column densites are large, ranging from $10^{21.5}~{\rm cm^{-2}}$ to $10^{23}~{\rm cm^{-2}}$ given the range of densities we consider.
The large column density facilitates Ly$\alpha$ trapping, which is another important mechanism to produce excited hydrogen at $n=2$ \citep{ferland1979,hall2007}.
The physical scale of the absorber can be approximated as $\Delta l = \frac{N_{\rm H}}{n_{\rm H}\epsilon}$, where $\epsilon$ is the gas volume filling factor.
If the absorber has $\epsilon \sim 1$, the physical scale implied by the matched models would be $\Delta l \lesssim 10^{-3}~{\rm pc}$, compatible with a single cloud within the BLR.

We note that, for a Galactic gas-to-dust ratio, the very minimum column density of $N_{\rm H}>10^{21.5}~{\rm cm^{-2}}$ inferred for the absorbing clouds, would imply a minimum dust absorption of $A_V>2.2$. This is inconsistent with the reddening inferred from the narrow lines, and also essentially inconsistent with the upper limit inferred from the broad lines. This is not a novel finding, as in AGN it has been known for a long time that dust absorption is much lower than what would be inferred from the absorbing gaseous column density \citep{Maiolino2001dust}. One of the primary interpretations of this discrepancy is that much of the gaseous column density (as typically inferred from the X-rays) is associated with (dust-free) BLR clouds \citep{Risaliti2002Xrayvar1,Risaliti2005Xrayvar2,Maiolino2010X-ray}. This perfectly fits the scenario that is emerging for GN-28074, in which we are finding evidence for absorbing clouds with density typical of the BLR clouds. As we will see below, these are also estimated to be within the sublimation radius.


\subsubsection{Location of the absorber}

The inferred gas densities are consistent with those of the BLR clouds, or clouds close to the BLR \citep{Netzer1990}.

In principle,
from the narrow range in $\log U$ derived for the absorber, and from the bolometric luminosity derived in \autoref{subsec:BH_prop} from the broad component of H$\alpha$, we can derive the distance of the absorber from the accretion disk.
Assuming the absorbing clouds are embedded in tenuous and optically thin gas \citep[i.e., a two-phase BLR model;][]{krolik1981}, and assuming the ionizing fraction and average ionizing photon energy from the \textsc{Cloudy} SED (which has a representative optical-to-X-ray slope of $\alpha _{\rm ox} = -1.4$), we use
\begin{equation}
    r_{\rm abs} = \sqrt{\frac{L_{\rm bol}f_0}{4\uppi c n_{\rm H}U \langle h\nu _0 \rangle}}
\end{equation}
to estimate the location of the absorber.
In this equation $f_0$ is the ionizing fraction and $\langle h\nu _0\rangle$ is the average ionizing photon energy.
We find the distance between the absorber and the source of the ionizing continuum to be from 1.9 pc to 0.19 pc, for $n_{\rm H} = 10^8~{\rm cm^{-3}}$ to $n_{\rm H} = 10^{10}~{\rm cm^{-3}}$ (this is without dust correction, but this and the following estimates of the sizes all scale as $L_{\rm bol}^{0.5}$, and thus the relative locations are not affected by dust correction). These sizes (although extremely uncertain, given the several assumptions) are larger than the radius of the BLR estimated by using the scaling relation from \cite{Greene2005blr}, which would give $r_{\rm BLR} \sim 0.036$ pc. Inferring the BLR size from the \Hbs luminosity scaling relations of \cite{Kaspi2005} gives a consistent radius of 0.04~pc.

We can also compare the estimated radial location of the absorber with the expected size of the dusty torus.
Assuming a dust sublimation temperature of $T_{\rm sub.}\approx 1,500$ K, the thermal equilibrium condition gives the inner radius of the torus to be $r_{\rm torus} = \sqrt{\frac{L_{\rm bol}}{4 \uppi \sigma T_{\rm sub.}^4}} \sim 0.21$ pc without dust correction.
Therefore, the location of the absorbing cloud can be between the BLR and the torus.
Interestingly, this inferred location is consistent with previous interpretations of Balmer-line absorbing AGN at lower redshifts \citep[e.g.,][]{hutchings2002,Zhang2015}, despite the fact that our derived column densities for GN-28074 are higher.
However, we recall that these estimations, especially the location of the clouds based on the ionization parameter, are extremely uncertain. 
This is because we have assumed optically thin conditions for clouds inner to the absorber to translate the ionization parameter to the distance, which ignores the potential self-shielding of the continuum emission in the BLR \citep{gaskell2009}.
It is still entirely possible that the absorbing clouds are actually associated with the outer parts of the BLR.

One might wonder what is the physical state of the absorbing gas. In our \textsc{Cloudy} simulations, the absorber has a warm temperature of $10,000\sim 20,000$ K and is nearly isobaric, which places it in a thermally stable regime \citep{reynolds1995}.

To check the dynamical state of the absorber, we compare the integrated radiation pressure from the attenuated continuum emission with the gravitational pressure.
The radiation pressure is from our \textsc{Cloudy} models, integrated until the depths that match the derived column densities.
The gravitational force induced solely by the central black hole is
\begin{equation}
    F_{\rm grav} = {4\uppi C_fGM_{\rm BH}\mu m_{\rm p} N_{\rm H}},
\end{equation}
where $\mu = 1.4$ is the mean atomic mass number and $m_{\rm p}$ is the proton mass.
The ratio between the radiation pressure and the gravitational pressure is then given by
\begin{equation}
    \frac{P_{\rm rad}}{P_{\rm grav}} = P_{\rm rad}/\left(\frac{GM_{\rm BH}\mu m_{\rm p} N_{\rm H}}{r_{\rm abs}^2}\right),
\end{equation}
which lies in a range of $0.72 - 25$.
Despite the large column density, the integrated radiation pressure upon the dense gas due to the attenuated continuum emission is comparable and even larger than the gravitational pressure, consistent with a radiatively driven outflow scenario.
The radiation pressure is only lower than the gravitational pressure when the column density of the absorber reaches $N_{\rm H} \sim 10^{23}~{\rm cm^{-2}}$.
Still, we caution that this estimate does not take into account the possibility that the continuum emission is partially shielded by BLR clouds inner to the absorber, which would in principle reduce the radiation pressure.

As we have discussed, the outflow velocity traced by \HeI absorption is larger than that traced by Balmer absorption.
If we assume the difference is real despite the complexity from the resonant scattering of \HeI, this might imply a decelerating outflow as the Balmer absorption can preferentially occur in a more neutral and thus outer region. The deceleration could happen as the outflow sweeps up the surrounding gas.
This is in contrast to the outflowing Balmer-line and \HeI-line absorber in a nearby AGN, NGC 4151, where the outflow is instead accelerating outwards \citep{hutchings2002}.
The decelerating scenario could indicate that the outflow will eventually fall back to the BLR and thus be self-replenished, maintaining the high covering factor.
In fact, if one simply estimates the escape velocity for the absorber based on the distances derived from the ionization parameter, one obtains
\begin{equation}
    v_{\rm esc} = \sqrt{\frac{2GM_{\rm BH}}{r_{\rm abs}}}\approx 800 - 3000~{\rm km\, \mathrm{s}^{-1}},
\end{equation}
where the lowest $v_{\rm esc}$ corresponds to $n_{\rm H} = 10^8~{\rm cm^{-3}}$ and the highest $v_{\rm esc}$ corresponds to $n_{\rm H} = 10^{10}~{\rm cm^{-3}}$.
Comparing these values to the outflow velocities in \autoref{tab:abs_par} which lies within $600-1000$ km s$^{-1}$, it is indeed plausible that while the gas is outflowing at the time of observations, it will eventually stall and fall back.
In addition, considering a two-phase or multi-phase BLR model \citep[e.g.,][]{krolik1981,reynolds1995}, as the dense absorber propagates outwards, the ram pressure from the volume-filling and highly ionized phase of the gas would help to slow and trap the outflow.
The ram pressure would become important when the ambient gas has a number density of $n_{\rm H}\gtrsim 10^4~{\rm cm^{-3}}$, consistent with tenuous gas with temperatures close to the Compton temperature and in pressure equilibrium with the typical line-emitting clouds in the BLR.
Furthermore, the absorber may continuously sweep up the gas in the outward direction, leading to increasing mass, which further works to slows the outflow. This would, in principle, also lead to strengthened absorption, but the effect might not be significant over a length scale of the size of the BLR, if the ambient density is $n_{\rm H}\sim 10^4~{\rm cm^{-3}}$.

\subsubsection{Nuclear neutral outflow}

Finally, we investigate the properties of the nuclear outflow traced by the blueshifted absorption features.
Estimating the total mass of the outflowing gas is not entirely straightforward as the absorbing clouds do not necessarily trace the entirety of outflowing gas.

We then use the following equation derived for a spherically symmetric outflow with $C_f = 1$ to estimate the outflow rate:
\begin{equation}
    \dot{M}_{\rm out; BLR} = 4\uppi R\mu m_{\rm p} N_{\rm H} v_{\rm out},
\end{equation}
where $\mu = 1.4$ is the mean mass per proton and $N_{\rm H}$ is the hydrogen column density. 
We can take the size of the outflowing medium as estimated in the previous subsection, and a minimum column density of $N_{\rm H}>10^{21.5}~{\rm cm}^{-2}$, as inferred by our models.
We use both velocity values from \autoref{tab:abs_par} in conjunction. The inferred lower limit on the nuclear outflowing mass is $0.05$~M$_{\odot}$~yr$^{-1}$.

This is a very conservative lower limit, as we are assuming the lowest of the radius estimates above and the bare minimum of the possible column densities. However, it is interesting to note that, based on the luminosity of the broad H$\alpha$ line and assuming a BLR gas density of $10^9-10^{10}~{\rm cm}^{-3}$, the inferred mass of the BLR is about $10-100~M_{\odot}$. If the nuclear outflow is tracing the outer part of the BLR, it would imply that the BLR would be wiped away in a few hundred/thousand years. These are very short timescales, suggesting that the BLR clouds are continuously replenished \citep[as inferred by other independent evidence][]{Maiolino2010X-ray}, probably lifted from the accretion disc, or simply through the eventual falling back of the outflow gas. An alternative possibility is that they are the bloated atmospheres of nuclear giant stars \citep{Scoville_Norman_1988,Alexander_Netzer_1994}, in which case those stars would continuously replenish the ``clouds'' on long timescales.

For the decelerating nuclear outflow scenario, while the outflowing dense gas would be recycled at a later time, it is currently radiation-pressure dominated and it might expand and become rarefied, creating a warm ambient medium, with a temperature below the Compton temperature of $\sim 10^7$ K.
It has been postulated that the warm gas that produces X-ray absorption features in AGN could be a product of a rarefied dense outflow \citep{reynolds1995}.
The warm clouds could have a temperature around $10^5$-$10^6$ K and lie at a location within or just outside the BLR (similar to what we find for the Balmer-line absorber), as inferred from modeling of previous observations of X-ray absorption in AGN \citep{reynolds1995,reynolds1997,mathur1997,komossa1997,nicastro1999,Chakravorty2009,Laha21}. Thus, under the isobaric condition, they would have a density of $n_{\rm H}\gtrsim 10^6~{\rm cm^{-3}}$.
While not contributing to Balmer absorption, the warm clouds can contribute to the obscuration of the X-ray emission at the location of the absorber.
Furthermore, the stall radius of the outflow can be calculated from the outflow velocity and the location of the absorber, which has a minimum value around 0.3 pc.
If the warm medium extends to the stall radius, can easily build up a column density of $N_{\rm H; warm} \gtrsim 10^{24}~{\rm cm^{-2}}$ and becomes Compton-thick.
While more quantitative calculations are needed to verify such a scenario, it provides an interesting picture for maintaining the X-ray weakness.

\subsection{X-ray properties}

\begin{figure}

    \includegraphics[width=0.5\textwidth]{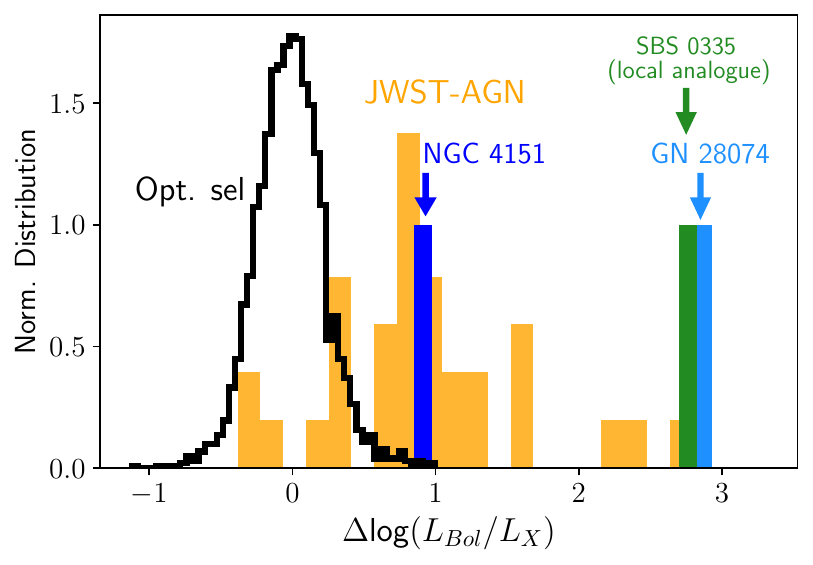}
    \caption{Distribution of the bolometric to X-ray luminosity ratio, relative to the relation inferred for low-redshift optically-selected AGN (whose distribution is shown with a black, hollow histogram). The distribution of JWST-discovered AGN in GOODS (the vast majority of which are actually lower limits on $L_{Bol}/L_X$) is shown with the solid, amber histogram. The lower limit for GN-280974 is indicated with the blue bar. We also show the lower limit observed in SBS 0335-052E, which is a nearby AGN with similar properties as the new population of the JWST-discovered AGN. We also show the value for the `prototypical' Seyfert 1 NGC-4151 which shows H$\alpha$, H$\beta$ and HeI absorption, just like GN-28074, and strong (and variable) X-ray absorption.}
    \label{fig:hist_Xrays}
\end{figure}

GN-28074 is located in a region of the Chandra Deep Field North (CDF-N) with excellent sensitivity within the context of the 2Ms Chandra survey \citep{Alexander2003}. Its X-ray analysis is reported in detail in \cite{Maiolino_xray_weak}, where they infer an upper limit of $< 2.9\times10^{41}$~erg~s$^{-1}$
on the 2-10~keV luminosity. When compared with the bolometric luminosity inferred from the broad lines, this implies a $L_{\rm bol}/L_{X} > 15400$, placing the object far above the standard $L_{\rm bol}$ - $L_{X}$ relation from \cite{Duras2020}, and actually making it the most X-ray weak among all AGN discovered by JWST and analysed in \cite{Maiolino_xray_weak}, with the measured upper limit being 3~dex lower than the expected X-ray luminosity of $\sim 10^{44}$~erg~s$^{-1}$.
Fig.\ref{fig:hist_Xrays} illustrates this more quantitatively by showing the distribution of the $L_{Bol}/L_X$ for JWST-selected AGN (amber histogram) relative to the low-z relation, for optically selected AGN (black hollow histogram), and keeping in mind that the values for the JWST-selected AGN are mostly lower limits on $L_{Bol}/L_X$. The latter are clearly offset towards much larger $L_{Bol}/L_X$ relative to low-z, optically-selected AGN, with GN-28074 being the most extreme of them all.

The X-ray weakness of GN-28074 is comparable to what seen in SBS~0335, a local metal poor dwarf galaxy, which was recently found to host an AGN, based on the detection of broad H$\alpha$ and variable mid-IR emission \citep{Hatano2023}, but which is extremely weak in the hard X-rays (the weak detection is probably associated with the star forming host).


Of the scenarios discussed in \cite{Maiolino_xray_weak} to explain the X-ray weaknesses of the JWST-discovered AGN, Compton thickness appears most likely, given the independent constraints on hydrogen column densities given by the absorption lines. However, we can not entirely discount intrinsic X-ray weakness, particularly given the extreme $L_{\rm bol}/L_X$ ratio, and additional information inferred from the radio emission discussed in the next section.

\subsection{Radio properties}
\label{subsec:radio_nondetection}

GN-28074 has been observed at 1.5~GHz, as part of the JVLA survey reported by \cite{Owen2018} and also part of the e-MERGE Survey \citep{Muxlow2020radio}. While the radio constraints on this and other JWST-discovered AGN will be discussed in more detail in a separate paper (Mazzolari et al. in prep.), here we simply report that the source is not detected at 1.5~GHz, with an upper limit of $1.6\times 10^{39}~{\rm erg\, \mathrm{s}^{-1}}$ on the luminosity at 1.5~GHz.

Based on the bolometric luminosity inferred from the broad lines, 
and using the bolometric-to-radio relationship for radio quiet AGN \citep{Bariuan2022radio,Wang2024radio}
one would expect a radio luminosity that is at least a factor between 5 and 20 larger than observed.

Once again, this radio weakness seems in line with the extreme radio weaknesses observed in SBS~0335 (Mazzolari et al. in prep.), which seems to be a local analogue of GN-28074 based on the X-ray and radio properties.

The radio weakness may be connected with the X-ray weakness in two possible ways.

The clouds responsible for the H$\alpha$ absorption are  also completely optically thick to free-free absorption. Indeed, the free-free absorption coefficient is given by

$$
\kappa _{ff} = 3.3\times 10^{-7}\left( \frac{n_{\rm e}}{{\rm cm}^{-3}}\right) ^2
\left( \frac{T_{\rm e}}{10^4~{\rm K}}\right)^{-1.35}
\left(\frac{\nu}{\rm GHz}\right)^{-2.1}~{\rm pc}^{-1}
$$
While the temperature of the clouds is always in the range between~ 5,000-25,000~K, the density of the clouds is so high that (despite their thickness of only $10^{-6}-10^{-3}~{\rm pc}$) the expected optical thickness for free-free should be of the order of $\tau _{ff} =  10^5-10^6$. Therefore, the radio emission should be completely absorbed by the clouds that absorb H$\alpha$. However, obviously the question is whether the radio emission (possibly associated with small jets, in radio quiet AGN), is compact enough to be entirely covered by such clouds. This remains an open question, however we note that in local AGN evidence for nuclear, sub-pc scale jets being free-free absorbed has been seen
\citep[e.g.][]{Taylor1996, Meyer2024}.

Another possibility is that the magnetic field in the nuclear region of these early AGN has not yet developed properly. Some 
recent zoom-in simulations do expect a weaker nuclear magnetic field in the early phases of galaxy formation
(S. Martin Alvarez, priv. comm.). A weak (or absent) nuclear magnetic field would obviously prevent significant radio emission as nuclear magnetic fields are believed to be a significant mechanism of launching AGN jets \citep{EHC2019, Zamaninasab2014}. At the same time, the hot corona responsible for the hard X-ray emission is thought to be produced by gas lifted from the accretion disc by the magnetic field \citep{Haardt1991, Cheng2020}. Therefore, a weak magnetic field would also automatically result in to an intrinsic X-ray weakness.

Lastly, it may be possible that this AGN has relatively recently launched a radio jet, which has weakened the corona resulting in X-ray weakness. Such a jet may be visible at MHz frequencies, however, followup radio observations are required to test this scenario.

\section{Ruling out alternative scenarios for the broad lines}
\label{subsec:alt_blr}

To explain the puzzling nature of JWST-confirmed AGN similar to GN-28074 that lack hard X-ray emission, it has been proposed recently that some of these sources are not real AGN but are SF-driven processes in disguise \citep[e.g.,][]{Yue2024X-ray,Kokubo2024}.
In this section, we discuss the possibility of alternative scenarios that create broad permitted lines.

\subsection{Ruling out hyperdense ultra-metal-poor galactic outflows}

The broad component of the permitted lines (H$\alpha$, H$\beta$, HeI, Pa$\beta$) does not have a counterpart in the forbidden lines and, in particular, in bright \OIII$\lambda$5007. This has always been considered as a clear evidence for emission from the BLR, where the densities are much higher than the critical density of \OIII$\lambda$5007.

However, it has recently been suggested that the broad permitted lines (H$\alpha$, H$\beta$, \HeI, Pa$\beta$) discovered by JWST might actually not be tracing the BLR of AGN but galactic outflows \citep{Yue2024X-ray,Kokubo2024}. In this subsection we discuss that this scenario is totally untenable, based on multiple lines of evidence, specifically for GN-28074 and, more broadly, for the several other cases discovered by JWST.

As already discussed, there is the indication of a galactic outflow component to {\it all} emission lines, with width intermediate between the narrow and the broad components of the permitted lines. However, here we are considering the more extreme scenario suggested by \cite{Yue2024X-ray} and \cite{Kokubo2024} that also the very broad component of H$\alpha$, H$\beta$, Pa$\beta$, and \HeI is also tracing a (much faster) galactic outflow.

In the outflow scenario there would be two possibilities for potentially suppressing the broad component of \OIII$\lambda$5007, which are an extremely low metallicity or a very high density. We discuss in the following that both scenarios are not possible.

As it is a strong coolant of the ISM, the intensity of the \OIII$\lambda$5007, relative to the Balmer lines, actually increases as the metallicity decreases, down to about 0.15-0.2~Z$_{\odot}$ \citep{Maiolino_mzr_2008,Curti_metcal_2017}, which is about the gas metallicity of GN-28074, as measured from our spectra and discussed in Appendix~\ref{appendix:met_ion_outflow}. Therefore, in the outflow scenario, the broad component of \OIII$\lambda$5007 should be even more prominent than H$\alpha$ and H$\beta$, contrary to what observed. As already discussed in \cite{Maiolino_xray_weak}, 
in order for the putative broad component of \OIII$\lambda$5007
to become significantly fainter than the corresponding broad component of the Balmer lines, the metallicity of the outflowing gas should be less than 0.01~Z$_\odot$, which is more than one order of magnitude lower than the host galaxy generating the outflow. This is the opposite of what seen in all galactic outflows, in which the metallicity of the outflowing gas is higher than the host galaxy.

Alternatively, the density of the outflowing gas should be much higher than the critical density of \OIII$\lambda 5007$. Specifically, in order not to be detectable the gas density should be higher than $10^7 ~{\rm cm}^{-3}$. With such high density, the
implied column of gas in the outflow, assuming an extension of about 10--100~pc
(i.e., about the size of the stellar component that should be driving the
outflow)
would result into a column of gas of about $10^{27}~{\rm cm}^{-2}$. For a dust-to-gas
ratio expected at a metallicity of 0.15~Z$_\odot$ \citep[see, e.g.,
][]{Deugenio2024}, would imply a dust attenuation of $A_V\sim 2 \times 10^4$.
This would totally absorb the receding side of the outflow, leaving visible only the
blueshifted side, which is completely in contrast with the symmetric profile of
all broad components. For the records, in galactic ouflows it only needs
$A_V\sim 1$ to result into a prominently blueshifted profile of the lines,
especially in H$\beta$. In order to comply with the hyper-dense outflow
scenario, the symmetric profiles of the broad lines would require essentially
no dust in the outflowing gas, or, equivalently, a metallicity of $\sim
10^{-4}~Z_\odot$, coming out of the nucleus of a galaxy with a metallicity of
$0.1~Z_\odot$.

The additional piece of evidence against the hyper-dense, ultra-metal poor
outflow 
scenario is the luminosity of the broad lines. If ascribed to an outflow, then
the luminosity of the broad component of H$\alpha$ should be associated with
photoionization by star formation. Even in the most conservative extinction
scenario, the H$\alpha$ luminosity would imply an
SFR$\sim$270~$M_\odot~\mathrm{yr}^{-1}$ \citep{Pflamm2007}. Such a huge SFR would have massive observable
implications at other wavelengths. It should be a bright submm source \citep[while
this galaxy is undetected in deep SCUBA2 surveys of GOODS-N, ][]{Geach2017} and it should have a
24$\mu$m flux at least one order of magnitude higher than observed.
Additionally, the X-ray emission associated with SNe and X-ray binaries 
should have a luminosity higher than $1.4\times 10^{42}~{\rm erg~s}^{-1}$, while the
upper limit from the non-detection in the Chandra 2Ms observation is about an
order of magnitude lower. Finally, given the small size of the source, the implied
surface density of SFR would be about $3\times 10^4~{\rm M_\odot~yr^{-1}~kpc^{-2}}$; such humongous star formation rate density has never been observed in any system and is orders of mangnitude higher than the theoretical maximum startburst limit of $10^3~{\rm M_\odot~ yr^{-1}~kpc^{-2}}$, given by the Eddington limited starbursts \citep{Thompson2009}.

Similar considerations apply for many other galaxies with broad H$\alpha$ discovered by JWST.

\subsection{Ruling out Raman scattering by neutral hydrogen}


\begin{figure}

    \includegraphics[width=0.48\textwidth]{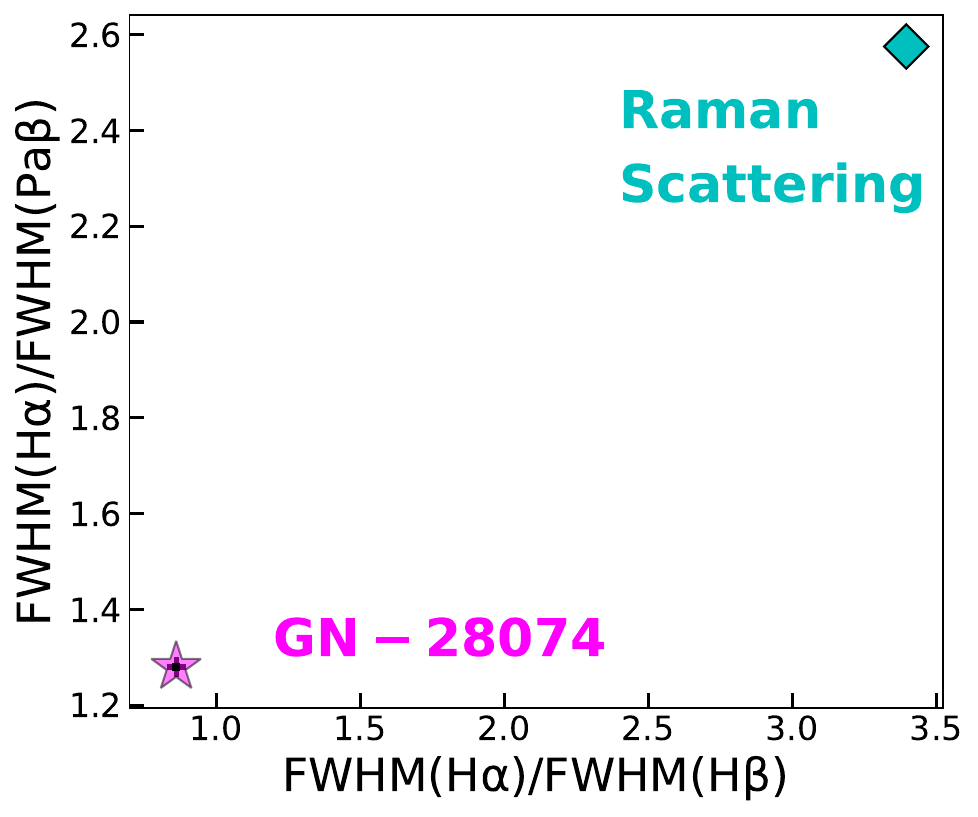}
    \caption{Comparison between the FWHM ratios of the broad hydrogen lines measured in the NIRSpec R1000 spectrum of GN-28074 and those predicted by a single-cloud inelastic Raman scattering model by \citet{kokubo2024a}. Clearly, the profiles of the broad hydrogen lines measured in GN-28074 cannot be explained by the Raman scattering of the UV continuum by neutral hydrogen.}
    \label{fig:raman_comparison}
\end{figure}

Another non-AGN scenario that can produce broad permitted emission in X-ray weak sources, as suggested recently by \citet{Kokubo2024}, is the inelastic Raman scattering of continuum emission bluewards of Ly$\alpha$ by neutral hydrogen at the ground state, where the continuum emission does not need to have an AGN origin.
This phenomenon was initially discussed to explain the emission observed in symbiotic stars \citep{Schimid_raman_1990,Schmid_raman_1994,Schmid_raman_1992}.
Since the Raman scattering cross-sections are relatively high around the wavelengths of Lyman series with broad wings, it can lead to broad absorptions around Lyman lines and subsequently broad emission around Balmer lines and other hydrogen recombination lines remembling a BLR component.

Significant Raman scattering requires the existence of neutral hydrogen at the $1s$ state with $N_{\rm H} \gtrsim 10^{21}~{\rm cm}^{-2}$, which is possible for the case of GN-28074 given the presence of the partially neutral absorber. However, as shown by \citet{kokubo2024a}, Raman scattering tends to create different line profiles for different hydrogen lines, with $\rm FWHM(H\alpha) \approx 3.4~FWHM(H\beta) \approx 2.6~FWHM(Pa\beta)$ for a single-cloud model (see Equations 32a-e in \citealp{kokubo2024a}, assuming the FWHM scales with the velocity width of the line).
In \autoref{fig:raman_comparison} we compare the predicted line widths from a single-cloud Raman scattering model with those measured in the R1000 spectrum of GN-28074.
While the FWHMs of hydrogen lines increase with increasing column densities under Raman scattering, the FWHM ratios are roughly constant, which means the comparison we make in \autoref{fig:raman_comparison} can be used to test Raman scattering in a wide range of cloud depths.
Based on our measurements for the broad emission-line component in GN-28074, $\rm FWHM(H\alpha)_{\rm broad} = 3610 \pm 22$ km s$^{-1}$ and $\rm FWHM(Pa\beta)_{\rm broad} = 2820 \pm 31$ km s$^{-1}$ (see \autoref{tab:kinematics}). These give $\rm FWHM(H\alpha)_{\rm broad}/FWHM(Pa\beta)_{\rm broad} = 1.28\pm 0.02 \ll 2.6$, which is significantly lower than the prediction from the Raman scattering.
We note that while the measured width of the broad \Ha\ is indeed larger than that of the broad Pa$\beta$, such a level of difference is not unexpected for a BLR origin based on previous observations of AGN \citep{landt_pabawidth_2008,lamperti_pabawidth_2017}.
As suggested by \citet{kim_pabaregion_2010}, the narrower Paschen lines could be explained by the scenario where they are preferentially emitted by clouds on the outskirts of the BLR compared to Balmer lines. 
In addition, the broad \Has and \Hbs lines are well fitted with the same width profiles, thus $\rm FWHM(H\alpha)_{\rm broad}/FWHM(H\beta)_{\rm broad} \approx 1 \ll 3.4$. When fitting their kinematics separately, $\rm FWHM(H\alpha)_{\rm broad} = 3600_{-23}^{+22}$~km~s$^{-1}$ while $\rm FWHM(H\beta)_{\rm broad} = 4180_{-160}^{+150}$~km~s$^{-1}$, giving a ratio of $0.86 \pm 0.03$, which is again inconsistent with Raman scattering. It should be noted here that, while the difference in line widths is significant ($\sim$3.5$\sigma$), separating the kinematics of \Has and \Hbs 
results in a worse fit in the outflow component for \OIIIs as well as creating significant residuals around the locations of Balmer absorptions.
Thus, we prefer to keep the kinematics tied.

Also, the broad component under Raman scattering would be tilted towards the red wing especially for Balmer lines \citep{kokubo2024a}, while the broad lines observed in GN-28074 are symmetric. The flux ratios between the broad lines would significantly deviate from Case B values, although this could be hidden by dust attenuation. Furthermore, such a high column density of gas to create significant Raman scattering ($N_{\rm H} \gtrsim 10^{21}~{\rm cm^{-2}}$) needs to be nearly dustless, otherwise the UV continuum needed for the inelastic scattering would be easily attenuated by dust. Clearly, there is dust in GN-28074 as indicated by the ratios of narrow lines.

Finally, the Raman scattering cannot explain the broad component in the \HeI$\lambda$10830 line as well as the broadening of the O\,{\sc i}$\lambda$8446 line (see \autoref{fig:CaT_NaD} in Appendix~\ref{appendix:other_lines}). Therefore, we conclude the broad components in the spectrum of GN-28074 cannot be produced by the Raman scattering and the BLR explanation is still preferred.

Similar considerations apply for other galaxies with detections of broad H$\alpha$ and other broad Balmer and Paschen lines discovered by JWST, although a careful comparison will be needed due to the weakness of broad lines other than H$\alpha$ in these low-luminosity AGN.

\section{Fraction of JWST-discovered AGN with Balmer absorption}
\label{subsec:blr_absorption}

As mentioned in the introduction, H$\alpha$ absorption has been detected in many of the broad-line AGN discovered by JWST \citep{Matthee2024,Kocevski2024,Wang2024_z7,Maiolino23c}. It is difficult to assess the fraction, due to the different selection methods and observing modes. However, as already pointed out by \cite{Maiolino_xray_weak}, the fraction of broad line AGN with H$\alpha$ absorption in each of those surveys seems to be about 10\%.

However, we argue that the true fraction is likely much higher still. Indeed, in order to see H$\alpha$ in absorption with JWST/NIRSpec, low-resolution prism spectra (which are the most commonly used) are totally inadequate, and also the medium resolution grating (R$\sim$1000) might not have high enough resolution if the absorption features are less than a few hundred km s$^{-1}$ in width. The high resolution gratings (R$\sim$2700) are in principle more appropriate, but this is the least used observing mode (primarily because spectra are truncated when this grating is used with the MSA); moreover, in many cases the resolution may come at expense of signal-to-noise ratio and only the brightest targets may allow the detection of absorption features. Finally, the absorption of H$\alpha$ may be detectable only if associated with some significant blueshift (or redshift), while, if close to the systemic redshift, it would be difficult to disentangle from the narrow component of H$\alpha$. Indeed, the JADES survey is starting to reveal some cases of H$\alpha$ in absorption with low velocity offset that is detectable simply because it is somewhat broader than the narrow component (D'Eugenio, priv. comm.). Therefore, it is likely that a much larger fraction of the type 1 AGN discovered by JWST have intrinsic H$\alpha$ absorption, but it is not detected because of the inadequate observing mode used, insufficient S/N, or because the absorption is hidden by the emission features.

Summarising, the fraction of broad line AGN with H$\alpha$ absorption is likely much larger than 10\%, among the JWST-discovered AGN. This is
much higher than found in previous surveys ($\sim 0.1$\%), such as the SDSS at low redshift. This may fit the scenario suggested by \cite{Maiolino_xray_weak} that the intermediate/low luminosity AGN discovered by JWST at high redshift may be characterised by a large covering factor of the BLR clouds, whose (dust-free/dust-poor) absorption in the X-rays may be responsible for their observed X-ray weakness.

\section{Local analogues}
\label{subsec:analogues}
It has been pointed out that high-z AGN discovered by JWST are peculiar in being extremely X-ray weak \citep{Maiolino_xray_weak} and probably also radio weak (Mazzolari et al. in prep.), relative to the low redshift, optically selected (and, obviously, X-ray and radio selected AGN). In these terms, GN-28074 is among the most extreme JWST-discovered AGN, in the sense that it is clearly a luminous type 1 AGN with luminous broad permitted lines (H and \HeI), strong mid-IR excess, but undetected in the X-ray and radio. Yet it is important to explore whether this prototype of JWST-discovered AGN has analogues in the local Universe, even if they may be rare  (or possibly, overlooked).

To begin with, it is interesting that what is considered to be the type 1 AGN `prototype', NGC~4151, is actually characterised by blueshifted H$\alpha$, H$\beta$ and \HeI absorption, just as in 
GN-28074. In NGC~4151 there is clear evidence for time variable absorption in the X-ray; while the X-ray absorption does not reach the Compton thick values inferred for GN-28074, it does reach a few times $10^{23}~{\rm cm}^{-2}$, which can absorb substantially the hard X-rays
\citep{Zoghbi2019}. More specifically, in the high-absorption regime, the observed $L_{\rm bol}/L_X$ ratio of NGC~4151 is in the regime of the JWST-discovered AGN at high-z (Fig.\ref{fig:hist_Xrays}), although not as extreme as GN-28074.

More compelling is the case of SBS0335-052E. This is a low metallicity (1/40~Z$_\odot$) dwarf galaxy that, until recently was thought not to host an AGN. However, it was recently found to host an AGN based on both the detection of broad permitted emission lines and variability in the mid-IR \citep{Hatano2023}. Interestingly, the AGN is undetected in the X-ray (the weak detection is ascribed to the host galaxy, hence used as an upper limit) and placing it at the same level of $L_{\rm obs}/L_X$ as GN-28074, as illustrated in Fig.\ref{fig:hist_Xrays}. In addition, SBS0335-052E is also undetected in the radio at 1.5 GHz (once again the weak detection is likely associated with the host galaxy and used as upper limit), and with a radio luminosity much weaker than expected even for radio a quiet SED. This will be discussed further in Mazzolari et al. (in prep.).
The similarity between GN-28074 and
SBS0335-052E in terms of broad emission lines, mid-IR excess, and X-ray and radio weakness, suggests similarities and connections between the AGN discovered by JWST with low metallicity dwarf galaxies.
Indeed, more of these dwarf galaxies have been found to host AGN based on the detection of both high ionization lines and a broad component of the H$\alpha$  (without forbidden counterpart), while being undetected in the X-rays \citep{Hatano2024Empress,simmonds2016}. Most interestingly, some of these show evidence for H$\alpha$ absorption \citep{burke2021}, just as in GN-28074, suggesting that also in (at least some of) these cases the X-ray weaknesses may be due to a dust-poor, Compton thick absorber.

More recently, \cite{Arcodia2024} used eROSITA to explore the X-ray emission of AGN identified in dwarf galaxies identified via nuclear variability in the UV-optical-IR. Although they detect 10\% of them, the remaining are undetected even in the stacking, just like the broad (and narrow) line JWST-discovered AGN 
at high redshift \citep{Maiolino_xray_weak}. They suggest that these light AGN (hosting intermediate mass black holes) may be missing the canonical X-ray corona and/or undergoing unusual accretion modes resulting in non-standard spectral energy distributions.

\section{Discussion and Conclusion}
\label{sec:conclusion}

GN-28074 can be considered an optimal case study to explore the nature and properties of the new population of AGN that is being discovered by JWST at high redshift. Indeed, it shares many of their features and, being among the most luminous and at a lower redshift, it allows to characterize features that are more difficult to investigate in the more distant and fainter population.

GN-28074 has prominent permitted broad lines (H$\alpha$, H$\beta$, Pa$\beta$, \HeI) without a counterpart in the forbidden lines (primarily \OIII), which clearly indicate that these originate from the Broad Line Region of a type 1 AGN with luminosity $L_{\rm bol}\sim 6\times 10^{45}~\mathrm{erg}\, \mathrm{s}^{-1}$.
Its SED also has a clear mid-IR excess tracing hot dust emission typical dusty torii surrounding AGN.

As with the vast majority of the AGN newly discovered by JWST, GN-28074 is not detected in the X-rays and, it is actually the most extreme of them all in terms of having the highest $L_{Bol}/L_X$ with a lower limit of 15,000. We note that many of the other JWST-selected AGN may have
similarly high bolometric to X-ray ratios, however in the case of GN-28074 we can set a more stringent lower limit because of its lower redshift and higher luminosity.

The NIRSpec spectrum reveals additional insights into this source. In particular, the detection of absorption of H$\alpha$, H$\beta$ and \HeI$\lambda$10830, blueshifted by a few 100 km s$^{-1}$. Given that H$\alpha$ and H$\beta$ are not resonant lines and that the $n=2$ level is very short lived, observing these two transitions in absorption requires very high densities in order to properly populate the $n=2$ level. More specifically, we have shown that, in order to properly reproduce the absorption seen in  H$\alpha$, H$\beta$ and \HeI, the absorbing medium must have a density of at least $10^8~{\rm cm}^{-3}$, which is well above the values inferred for the ISM, while typical of the Broad Line Region (BLR) of AGN. The very small inferred sizes of the absorber ($10^{-6}-10^{-4}$ pc) are also consistent with those expected for the BLR clouds (although the observed absorption is likely the superposition of multiple clouds). The inferred {\it minimum} column density of the absorbing gas is $10^{21.5}-10^{23}~{\rm cm}^{-2}$, but the total column density of the multiphase gas is likely much larger. These findings support the scenario in which the extreme X-ray weakness is due (at least partly) to heavy absorption by Compton thick gas (dust-poor or dust-free), likely associated with the BLR clouds. The latter are known to be dust free and can therefore explain the widespread X-ray weakness of broad-lined, type 1 AGN discovered by JWST.

Actually, although with a large uncertainty, we infer that the location of the absorbing medium is between the BLR and the dusty torus. Therefore, H$\alpha$, H$\beta$ and \HeI absorption may arise in dense neutral gas clouds in the outer part of the BLR and in the process of outflowing. However, we shall note that these outflow velocities are modest (a few hundred km s$^{-1}$), much slower than the velocities seen in the nuclear region of quasars (e.g., in the case of BAL quasars), characterized by velocities of several thousands km s$^{-1}$ in the optical lines and up to $10^4-10^5$ km s$^{-1}$ in X-ray spectra.


We have discussed that, because of observational difficulties in detecting H$\alpha$ absorption, the fraction of broad-line AGN at high-$z$ characterized by this feature is likely much larger than the $\sim$10\% currently estimated (which is already much larger than the fraction estimated for low-$z$ AGN obtained in previous surveys). This may indicate that the population of intermediate/low luminosity AGN discovered by JWST at high redshift is characterized by a larger covering factor of dense (and dust-poor/free) BLR clouds, than lower redshift counterpart. The associated (Compton-thick) absorption may explain, or contribute to their X-ray weakness.

Another interesting feature of GN-28074 is its radio weakness. Once again, this seems to be a feature common to many other JWST-discovered AGN (Mazzolari et al., in prep.), but in GN-28074 it is most tightly constrained because of its high bolometric luminosity. The fact that it is radio weaker than radio quiet quasars (which are X-ray ``normal'') suggest that the radio weakness and X-ray weakness are somehow connected in this class of AGN. We have discussed that one possibility is free-free absorption. Indeed, the clouds responsible for H$\alpha$ absorption are certainly optically thick in free-free absorption; in this case the absence of radio emission would be associated with the X-ray Compton thick absorption. The only concern is whether the extent of the distribution of absorbing clouds is large enough, and with a large enough covering factor, to absorb the radio emitting region (presumably a small jet) -- however, direct observational evidence of jets absorbed by a free-free thick nuclear medium has been observed.
The other possibility is that the nuclear magnetic field has not yet developed properly: this would both prevent the radio emission (absence of a radio jet) and would hamper the lifting of gas from the accretion disc to produce a hot corona, which is required for the hard X-ray emission.

While it has been suggested that the broad emission line components in these X-ray weak (and Radio weak) sources could originate from non-AGN activities, using our measurements for GN-28074, we rule out this possibility.
Specifically, we explore alternative explanations for broad hydrogen lines including hyperdense or ultra metal-poor outflows and Raman scattering of the (stellar) UV continuum emission by a significant amount of neutral hydrogen.
For the ultra metal-poor outflow scenario, the outflow needs to be a least an order of magnitude more metal-poor than the ISM, which is the opposite of what usually observed in galactic outflows.
As for the hyperdense outflow scenario, the implied amount of dust attenuation would create significantly blueshifted profiles, in contrast to the symmetric broad profiles observed, unless the outflow is orders of magnitudes more metal-poor.
In addition, the galactic outflow scenario would require an extreme SFR surface density above the theoretical Eddington limit and make GN-28074 very luminous in the submm, which is not detected in either SCUBA2 or NOEMA.
The Raman scattering scenario, on the other hand, is also disfavored as the measured ratios of line widths of broad hydrogen lines are significantly different from the predictions.
Detections of broad components in \HeI$\lambda 10830$ and O\,{\sc i}$\lambda 8446$ further rule out the Raman scattering scenario.
As a result, the AGN origin for the broad permitted lines is still favored over other scenarios.

We finally note that local analogues of the AGN discovered by JWST at high-$z$ do exist. The fact that the, thoroughly studied, Seyfert 1 ``prototype'' NGC~4151 does have H$\alpha$, H$\beta$ and \HeI absorption, similar to GN-28074, and has substantial absorption in the X-rays, already indicates that GN-28074 (and likely many other JWST-discovered AGN) is not so peculiar. However, probably the closest and clearest example is SBS 0335-052E. This source has been traditionally considered to be an extremely metal-poor star-forming dwarf galaxy \citep{Izotov+1990}. Only recently observations have revealed the clear presence of an AGN via the detection of a broad component of H$\alpha$ (without a counterpart in the forbidden lines; \citealp{Hatano2023}), high ionization lines,  and strong mid-IR variability \citep{Hatano2023}.
While there are some differences (for instance, the SBS 0335-052E has a blue rest-optical continuum and a clear Balmer jump, whereas GN-28074 displays red optical continuum and a spectral break), there are also strong similarities.
Like GN-28074 (and other JWST-discovered AGN), SBS 0335-052E is extremely X-ray weak (and the little X-ray emission detected is likely associated with the star-forming host) and also does not show evidence of nuclear radio emission at 1.5~GHz (Mazzolari et al, in prep.). Additionally, recent studies have shown that more local low-metallicity dwarf galaxies host similar AGN, which have been missed because of the lack of X-ray and radio emission.

In summary, GN-28074 at intermediate redshift, and SBS 0335-052E locally, are important `Rosetta stones' to be investigated with additional observations and modelling to better understand the nature of the new, large population of AGN discovered by JWST.

\section*{Acknowledgements}
I.J. acknowledges support by the Huo Family Foundation through a P.C. Ho PhD Studentship. IJ, XJ, RM, FDE, JS, GM and JW acknowledge support by the Science and Technology Facilities Council (STFC), by the ERC through Advanced Grant 695671 “QUENCH”, and by the UKRI Frontier Research grant RISEandFALL. RM also acknowledges funding from a research professorship from the Royal Society. SA acknowledges grant PID2021-127718NB-I00 funded by the Spanish Ministry of Science and Innovation/State Agency of Research (MICIN/AEI/ 10.13039/501100011033). AJB acknowledges funding from the ``FirstGalaxies'' Advanced Grant from the European Research Council (ERC) under the European Union’s Horizon 2020 research and innovation programme (Grant agreement No. 789056). ECL acknowledges support of an STFC Webb Fellowship (ST/W001438/1). M.P. acknowledges support from the research project PID2021-127718NB-I00 of the Spanish Ministry of Science and Innovation/State Agency of Research (MICIN/AEI/ 10.13039/501100011033), and the Programa Atracci\'on de Talento de la Comunidad de Madrid via grant 2018-T2/TIC-11715. PGP-G acknowledges support from grant PID2022-139567NB-I00 funded by Spanish Ministerio de Ciencia e Innovaci\'on MCIN/AEI/10.13039/501100011033, FEDER, UE. BER acknowledges support from the NIRCam Science Team contract to the University of Arizona, NAS5-02015, and JWST Program 3215. ST acknowledges support by the Royal Society Research Grant G125142. H{\"U} gratefully acknowledges support by the Isaac Newton Trust and by the Kavli Foundation through a Newton-Kavli Junior Fellowship. JW acknowledges support from the Science and Technology Facilities Council (STFC), by the ERC through Advanced Grant 695671 “QUENCH”, by the UKRI Frontier Research grant RISEandFALL. The work of CCW is supported by NOIRLab, which is managed by the Association of Universities for Research in Astronomy (AURA) under a cooperative agreement with the National Science Foundation.

\section*{Data Availability}
All data used in this study are now public on the JWST DAWN archive.



\bibliographystyle{mnras}
\bibliography{example,roberto_GNz11} 




\appendix

\section{Calcium triplet, sodium absorption and fluorescent oxygen lines}
\label{appendix:other_lines}

\begin{figure*}
    \centering
    
    \subfloat{\includegraphics[width=1\columnwidth]{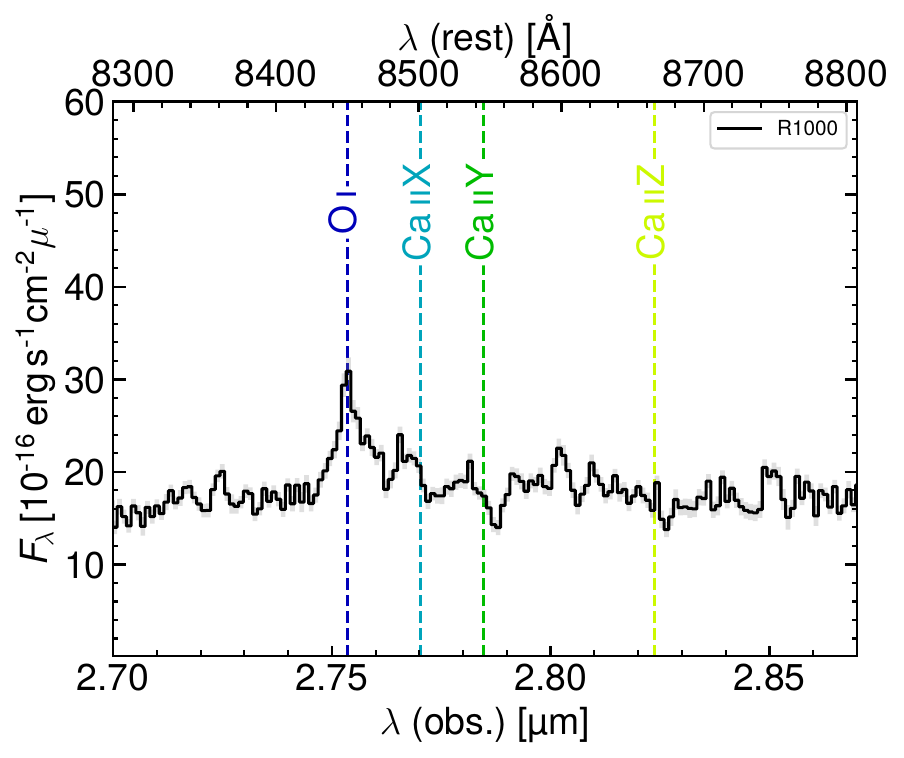}} \hfill
    \subfloat{\includegraphics[width=1\columnwidth]{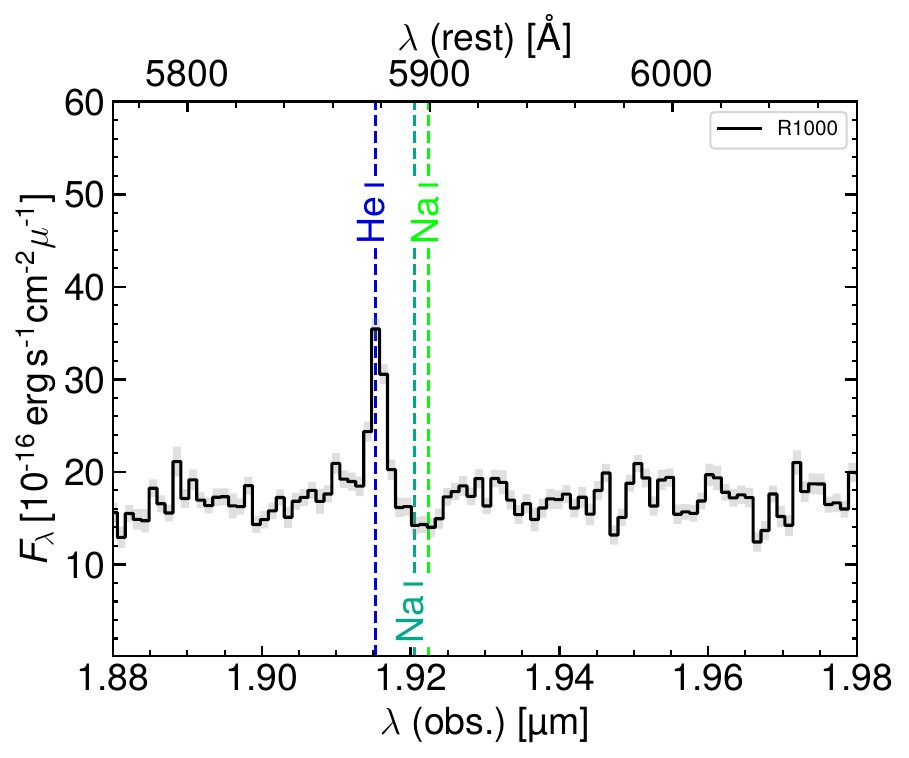}} \hfill
    \caption{Plots showcasing the positions of the calcium and sodium absorption features with respect to the OI$\lambda$8446 and \HeI$\lambda$5875 lines respectively. The observed spectrum is shown with the solid black line, errors are indicated by a shaded grey region.}
    \label{fig:CaT_NaD}
\end{figure*}

In addition to the clear absorption in the \Ha, \Hbs and \HeI$\lambda$10830 lines 28074 also exhibits CaT and NaD absorption, shown in \autoref{fig:CaT_NaD}.

\begin{figure}
    \centering
    \includegraphics[width=\columnwidth]{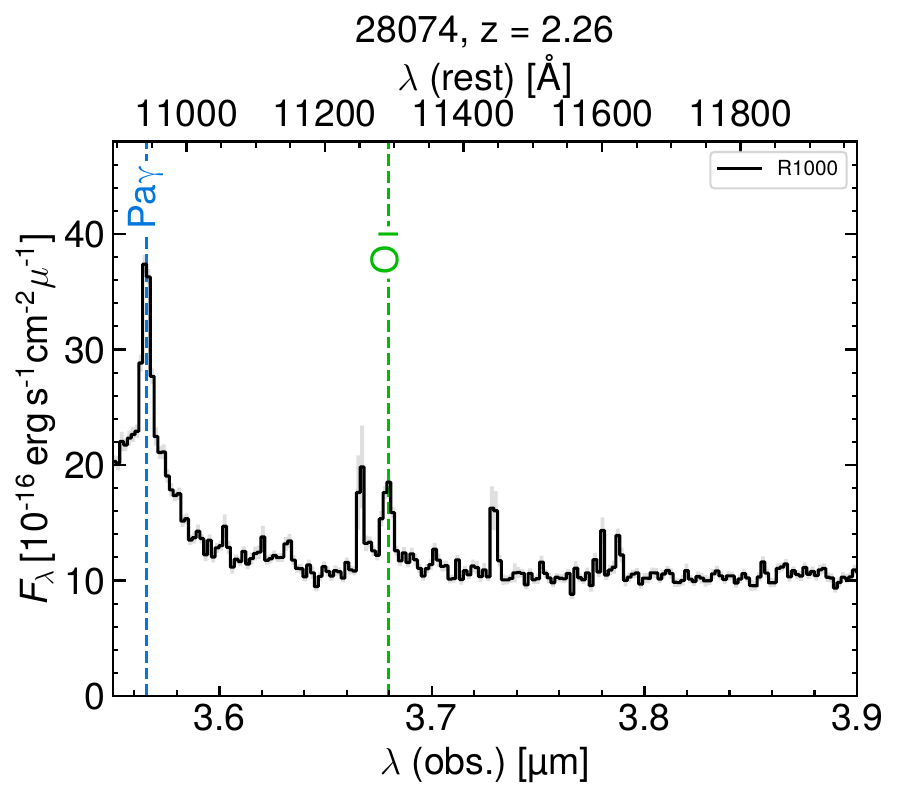}
    \caption{Region containing the O\,{\sc i}$\lambda$11287 line, the position of which is indicated by a dashed green line. The R1000 grating data is shown by the black line, 1$\sigma$ errors are indicated by a shaded green region. Sharp feature bluewards of the oxygen line is likely a data artefact as evidenced by the pipeline errors being considerably larger for that data.}
    \label{fig:oi_11287}
\end{figure}

As can be seen in the figure, the sodium doublet absorption appears to coincide with the systematic redshift of the narrow lines, while the calcium triplet appears redshifted by about 300~km~s$^{-1}$. As the triplet absorption is usually associated with stellar populations, a possible reason for the systematic redshift may be an underlying merger in the host galaxy. Following this interpretation, the reason for the lack of displacement of the sodium absorption may be resonant scattering in the underlying gas.

However, the kinematics of the calcium triplet region of the spectrum are difficult to precisely model due to underlying broad Ca\,{\sc ii} emission as well as the nearby O\,{\sc i}$\lambda$8446 line, while the sodium absorption is somewhat tentative (3-4$\sigma$ significance). Therefore, we can not draw any strong conclusions about the origin of these features.

The clearly detected O\,{\sc i}$\lambda$8446 and O\,{\sc i}$\lambda$11287 (\autoref{fig:oi_11287}) lines have a photon flux ratio of around unity indicating that they are likely produced by Ly$\beta$ fluorescense, as is common in AGN \citep{Rudy1989}. However, constraints on O\,{\sc i}$\lambda$1304 line are required to establish the Ly$\beta$ pumping scenario more robustly \citep{Rodri2002} while the O\,{\sc i}$\lambda$1304 line is not covered by the wavelength range of the NIRSpec spectrum of GN-28074.
Still, since there is no detection of O\,{\sc i}$\lambda$13165 and O\,{\sc i}$\lambda$7774 lines, the observed O\,{\sc i}$\lambda$8446 is unlikely to have significant contributions from recombination, collisional excitation, and continuum fluorescence \citep{grandi1980,Rodri2002}.
The Ly$\beta$ pumping scenario, on other hand, is consistent with the presence of dense gas optically thick to Balmer lines.
Recently, \citet{choe2024} reported detection of O\,{\sc i}$\lambda$8446 by JWST from an object in the gravitationally lensed Sunburst Arc at $z = 2.37$. According to \citet{choe2024}, the O\,{\sc i}$\lambda$8446 emitting object is likely an $\eta$ Carinae analog surrounded by dense gas, which produces O\,{\sc i}$\lambda$8446 by Ly$\beta$ fluoresence.
In the case of GN-28074 analysed in this paper, the dense gas that produces O\,{\sc i}$\lambda$8446 is likely associated with regions surrounding the accreting black hole due to the presence of very broad components in permitted lines that are unlikely to be produced by stars.

\section{Gas-phase metallicity}
\label{appendix:met_ion_outflow}

\begin{figure*}
    \subfloat[\OII $\lambda \lambda$3726,3729]{\includegraphics[width=1\columnwidth]{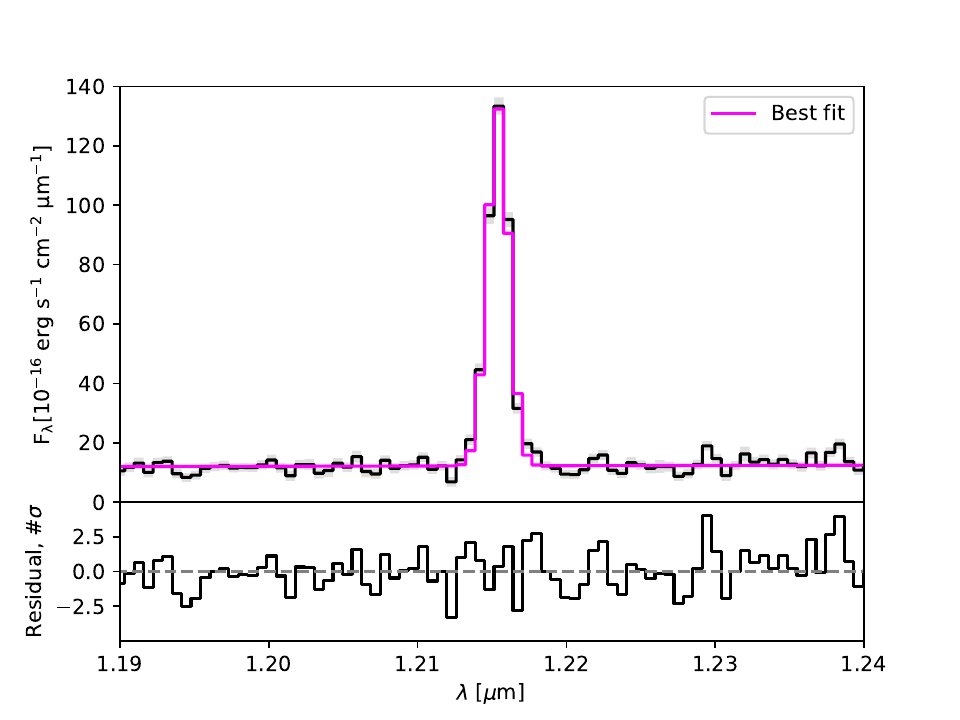}} \hfill
    \subfloat[\OII$\lambda \lambda$7320,7331]{\includegraphics[width=1\columnwidth]{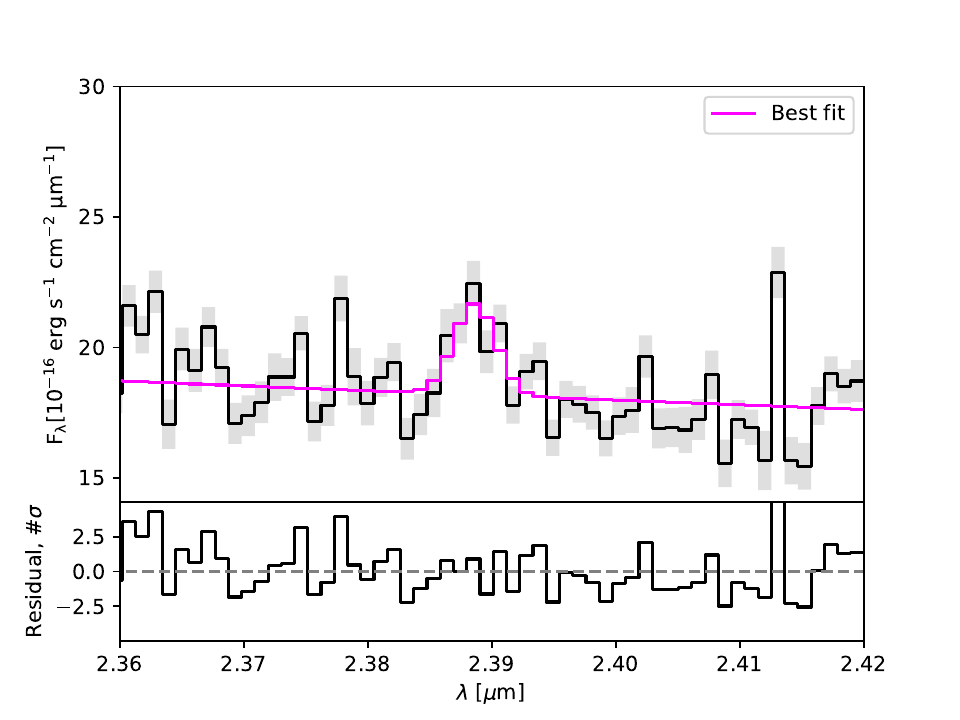}} \hfill
    \caption{Fits to the \OII$\lambda \lambda$3726,3729 doublet and its auroral counterpart. The \OII$\lambda \lambda$7320,7331 doublet appears marginal in R1000, despite having a formal significance of $\sim 6\sigma$.}
    \label{fig:auroral}
\end{figure*}

The clear presence of the \OIII$\lambda$4363 auroral line along with a tentative detection of the \OII$\lambda\lambda$7320,7331 doublet (\autoref{fig:auroral}) allows for robust estimates of $T_e$ in both $\rm O^+$ and $\rm O^{++}$ region. Utilizing the auroral to strong line ratios as well as an electron density of $n_e = 170_{-130}^{+232}$~cm$^{-3}$, inferred from an observed \SII$\lambda$6716/\SII$\lambda$6731 ratio of $1.32_{-0.13}^{+0.09}$, we estimate the $T_e({\rm O^{++}}) = 2.08_{-0.07}^{+0.06}\times10^{4}$~K and $T_e({\rm O^{+}}) = 1.17_{-0.15}^{+0.18}\times10^{4}$~K. 
Such a large difference in the temperatures of high-ionisation lines and low-ionisation lines are typical for the narrow line regions (NLRs) of AGN rather than star-forming regions \citep[e.g.,][]{garnett1992,perez-montero2014,Dors2020}. With these temperatures, we derive a metallicity of $\rm 12+\log(O/H) = 7.85_{-0.12}^{+0.17}$ or about 15\% Solar abundance when using the $A_V$ derived in \autoref{subsec:BH_prop}.

\section{Large scale ionized outflow}
\label{appendix:outflow}

\begin{figure}
    \centering
    \includegraphics[width=\columnwidth]{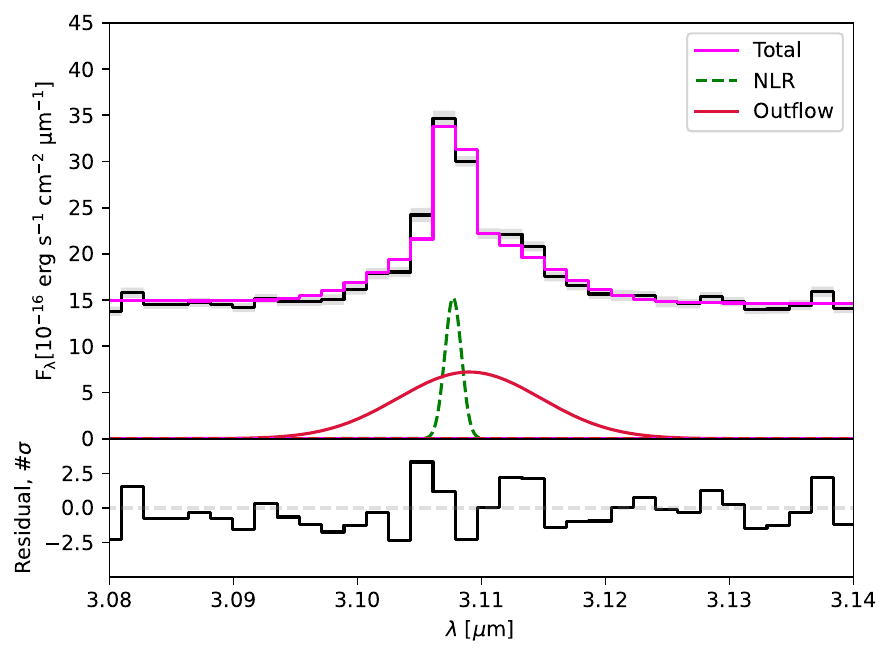}
    \caption{Best fit model of the \SIII$\lambda$9532 line emission. The line profile exhibits clear broad wings that are fitted well by an outflow component.}
    \label{fig:SIII_outflow}
\end{figure}

The fit of the \OIII doublet in the top left panel of \autoref{fig:spectrum_abs} clearly reveals the presence of weakly redshifted ionized outflow, seen also in the \SIII$\lambda$9531 line profile (\autoref{fig:SIII_outflow}). A component with matching kinematics is required to fully explain the shape of the \Has and \Hbs lines, without which the absorption profiles yield inconsistent estimates for the column density of hydrogen in the n = 2 state. However, an outflow component is disfavoured by fitting in the \NIIs and \SII\ doublets as well as the \HeI$\lambda$10830 and Pa$\gamma$ lines.

The fit to the outflow component reveals a FWHM~$=1310_{-19}^{+19}$~km~s$^{-1}$ and a slight, $\sim 130$~km~s$^{-1}$, redshift with respect to the narrow lines. We use the formalism of \cite{Carniani2015} to estimate the outflow mass traced by \OIIIs via:
\begin{equation}
    M_{\rm [OIII]} = 1.7\times 10^3 \frac{m_{\rm p}CL_{\rm [OIII]}}{10^{\rm [O/H] - [O/H]_{\odot}}j_{\rm [OIII]}n_{\rm e}},
\end{equation}
where $C = \langle n_{\rm e}\rangle^2/\langle n_{\rm e}^2\rangle$, $\rm [O/H] \equiv \log{(O/H)}$, $j_{\rm [OIII]}$ is the emissivity of the \OIIIs doublet and $L_{\rm [OIII]}$ - the luminosity of the outflow component in \OIII. Our fits to the \SII\ doublet allow us to estimate $n_{\rm e} \sim 100$~cm$^{-3}$. As we do not detect the broad component in the \SII\ emission lines, we assume that the outflow has the same density as the ISM; this may not be entirely representative as AGN driven outflows tend to be somewhat denser than the host ISM \citep{Davies2020, Santoro2020}. Under these density assumptions, along with the previous $T_e(O++) = 2\times 10^4$~K estimate to hold for the outflowing gas we estimate $j_{\rm [OIII]} = 1.2\times 10^{-20}$~erg~s$^{-1}$~cm$^{-3}$ using \textsc{PyNeb} \citep{PyNeb}. Using this emissivity along with the extinction corrected outflow luminosity yields $M_{\rm [OIII]} = 3.60^{+0.57}_{-0.46}\times 10^7$~M$_{\odot}$. It should be noted that the mass given by \OIIIs emission should be treated as a lower limit as this emission generally does not trace the entire outflow \citep{Carniani2015, Liu2013}. We estimated outflow velocity to be $1240_{-18}^{+18}$~km~s$^{-1}$ by using the same formalism as before - $v_{\rm out} = |\Delta v| + 2\sigma$. Assuming a spherical geometry the mass outflow rate can then be estimated as \citep{Maiolino2012}:
\begin{equation}
    \dot{M}_{out} = \frac{3v_{out}M_{out}}{R_{out}},
\end{equation}
where $v_{out}$ is the velocity, $M_{out}$ - the mass and $R_{out}$ - the radius of the outflow. Taking $R_{out} = 1$~kpc as an order of magnitude estimate based on the usual extent of AGN driven outflows \citep{Harrison2014, Carniani2015}, we obtain  $\dot{M}_{[OIII]} = 137_{-18}^{+21}$~M$_{\odot}$~yr$^{-1}$.

In addition to the \OIIIs doublet we also utilize the outflow emission detected in the \Hbs and \Has lines to estimate the mass and rate of outflowing hydrogen. We warn however, that these component are more uncertain than for the \OIIIs, as they are completely blended with the broad components of the same lines. Using the same derivation as in \cite{Carniani2015} we employ the following equation for the hydrogen outflow mass:
\begin{equation}
    M = 0.8\frac{m_pCL_{ik}}{j_{ik}n_e},
\end{equation}
where $L_{ik}$ and $j_{ik}$ are the luminosity and emissivity of a particular hydrogen transition respectively. Utilizing PyNeb as before we find $j_{H\beta} = 6.6\times10^{-26}$~erg~s$^{-1}$~cm$^{-3}$ and $j_{H\alpha} = 1.8\times10^{-25}$~erg~s$^{-1}$~cm$^{-3}$. Which lead to $M_{H\beta} = 13.0_{-2.4}^{+2.7}  \times 10^{7}$~M$_{\odot}$ and $M_{H\alpha} = 13.3_{-3.2}^{+4.2} \times 10^{7}$~M$_{\odot}$ - these estimates are consistent with each other and correspond to a total hydrogen outflow rate $\dot{M}_H = 500_{-110}^{+130}$~M$_{\odot}$~yr$^{-1}$, with the \OIIIs derived outflow rate providing a lower limit. We note that the Paschen line emissivity in this regime is an order of magnitude below that of \Hbs thus the outflow component there is likely buried beneath the more luminous BLR emission. On the other hand, the emissivity of \HeI$\lambda$10830 is estimated to be on the order of that of \Ha, however, the outflow emission there can be strongly affected by resonant scattering thus its absence is not very constraining. All ionized outflow properties are summarized in \autoref{tab:ion_of}.

\begin{table}
    \centering
    \renewcommand{\arraystretch}{1.3}
    \begin{tabular}{cccccc}
    \hline
        Line & $L_{\rm out}$ & $M_{\rm out}$& $v_{\rm out}$~[km~s$^{-1}$] & $\dot{M}_{\rm out}$ & P$_k$ \\
        \hline
         \OIIIs & $2.33_{-0.30}^{+0.37}$ & $3.60^{+0.57}_{-0.46}$ & \multirow{3}{*}{$1240_{-18}^{+18}$} & $137_{-18}^{+21}$& $6.85_{-0.97}^{+1.14}$ \\
         \Hbs & $1.27_{-0.23}^{+0.26}$ & $13.0_{-2.4}^{+2.7}$ & & $492_{-91}^{+103}$& \multirow{2}{*}{$24.2_{-5.2}^{+6.4}$}\\
         \Has & $3.6_{-0.9}^{+1.1}$ & $13.3_{-3.2}^{+4.2}$ & & $500_{-120}^{+160}$& \\
         \hline
    \end{tabular}
    \caption{A summary of the observed ionized outflow properties. Column one gives the name of the line, second column - the $A_V$ corrected outflow luminosity in units of 10$^{42}$~erg~s$^{-1}$, third column - the outflowing mass in units of 10$^{7}$~M$_{\odot}$, the fourth - outflow velocity in km~s$^{-1}$. Column five shows the mass outflow rates in M$_{\odot}$~yr$^{-1}$, while the final column gives the outflow kinetic power in units of $10^{43}$~erg~s$^{-1}$.}
    \label{tab:ion_of}
\end{table}

\begin{figure}
    \centering
    \includegraphics[width=\columnwidth]{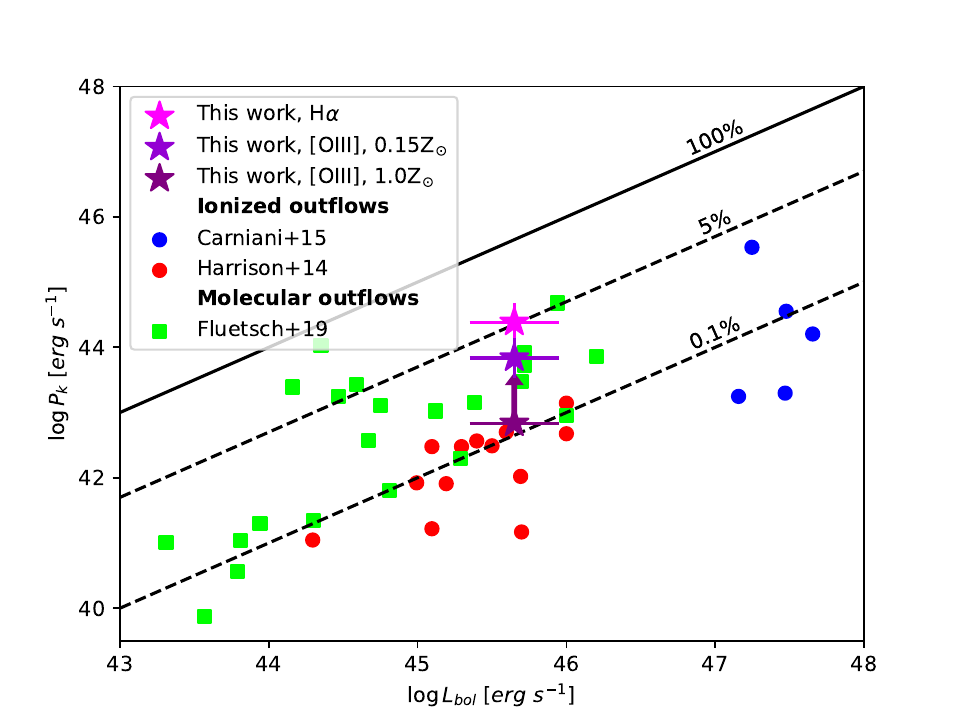}
    \caption{A comparison between the AGN bolometric luminosity, on the x axis, and outflow kinetic power, on the y axis, for 28074 and QSOs from \citet{Carniani2015}, \citet{Harrison2014} and \citet{Fluetsch2019}. Molecular outflow data is marked with green squares, while ionized outflows - with circles. Our object is indicated by magenta and violet stars for estimates using \Has and \OIIIs emission at NLR metallicity respectively. The lower limit obtained from \OIIIs emission assuming solar metallicity is marked by the purple star. The error bars indicate $\sim$0.3~dex systematic uncertainties. The literature data has been adjusted for the factor of 3 discrepancy between assumed geometries. The lines denote constant ratios of 100\%, 5\% and 0.1\%.}
    \label{fig:of_power}
\end{figure}

It is also possible to estimate the outflow kinetic power via the relation:
\begin{equation}
    P_k = \frac{1}{2}\dot{M}_{out}v_{out}^2,
\end{equation}
where $\dot{M}_{out}$ and $v_{out}^2$ are the outflow mass rate and velocity respectively. This equation yields $P_k = 2.42_{-0.52}^{+0.64}\times10^{44}$~erg~s$^{-1}$ when using the \Has emission and $P_k = 6.85_{-0.97}^{+1.14}\times10^{43}$~erg~s$^{-1}$ when using the \OIIIs doublet, corresponding to around 1-5\% of total AGN luminosity. This ratio is about 1-2~dex above the expected value for quasars at similar redshifts, as illustrated in \autoref{fig:of_power}, and is consistent with what found for molecular outflows. However, it should be noted that the \OIIIs outflow rate was calculated assuming the NLR metallicity for the outflowing gas, while outflows tend to be more metal rich than their hosts \citep{Chisholm2018}. Assuming Solar metallicity brings the outflow rate down by about 1~dex and results in a lower limit of $P_k = 6.85_{-0.97}^{+1.14}\times10^{42}$~erg~s$^{-1}$. In addition, the comparisons in \autoref{fig:of_power} are somewhat complicated by differing outflow velocity estimations used across the literature. In the case of data from \cite{Harrison2014}, the outflow velocities were taken to be FWHM/1.3 for single Gaussian profiles, which would lower our estimates on outflow power by about 50\%. To reflect these systematic differences, 0.3~dex was added to the error budget on the outflow kinetic power.

\bsp	
\label{lastpage}
\end{document}